\documentclass[reprint,amsmath,amssymb,aps,pra,floatfix]{revtex4-2}

\usepackage{graphicx}
\usepackage{dcolumn}
\usepackage{bm}
\usepackage{hyperref}

\usepackage[section]{placeins}

\usepackage[dvipsnames]{xcolor}

\hypersetup{
	colorlinks = true,
	linkcolor  = Blue,
	citecolor  = BlueViolet,
	urlcolor   = BlueViolet,
}

\usepackage{siunitx}

\graphicspath{{figures/}}

\newcommand{\ima}{\text{i}}
\newcommand{\dd}{\text{d}}

\DeclareMathOperator{\Arg}{Arg}
\newcommand{\Real}{\mathrm{Re}}
\newcommand{\Imag}{\mathrm{Im}}
\newcommand{\pol}[2][]{\text{e}^{{#1}\ima{#2}}}
\newcommand{\pdv}[3][]{\frac{\partial^{#1}{#2}}{\partial{#3}^{#1}}}
\newcommand{\abs}[1]{\left\lvert {#1} \right\rvert}
\newcommand{\expval}[1]{\langle {#1} \rangle}
\newcommand{\ket}[1]{\lvert {#1} \rangle}
\newcommand{\braket}[2]{\langle {#1} | {#2} \rangle}

\begin{document}
    \preprint{APS/123-QED}

    \title{Dynamics and fragmentation of bosons in an optical lattice inside a cavity using Wannier and position bases}%


    \author{Christopher Gerard R. Sevilla}%
    \email{csevilla@nip.upd.edu.ph}%
    \affiliation{National Institute of Physics, University of the Philippines, Diliman, 1101 Quezon City, Philippines}%

    \author{Jayson G. Cosme}%
    \email{jcosme@nip.upd.edu.ph}
    \affiliation{National Institute of Physics, University of the Philippines, Diliman, 1101 Quezon City, Philippines}%

    \date{\today}

    \begin{abstract}
        The atom-cavity system is a versatile platform for emulating light-matter systems and realizing dissipation-induced phases, such as limit cycles (LCs) and time crystals. Here, we study the dynamics of a Bose-Einstein condensate (BEC) inside an optical cavity with transverse pumping and an additional intracavity optical lattice along the cavity axis. Specifically, we explore the theoretical predictions obtained from expanding the atomic field operators of the second-quantized Hamiltonian in two ways: (i) position basis and (ii) single-band Wannier basis. Both bases agree on the existence of most types of static and dynamical phases. However, the large sea of irregular dynamical phase, captured within the position basis, is absent in the Wannier basis. Moreover, we show that they predict different types of LCs due to the inherent limitation of the single-band Wannier expansion, highlighting the importance of including higher energy bands to correctly capture certain phenomena. Using truncated Wigner approximation, we investigate the fragmentation dynamics of the BEC. We demonstrate that both position and Wannier bases qualitatively agree on the photon-mediated fragmentation dynamics of the BEC in the density-wave phase, despite the absence of interatomic interactions. The presence of interatomic interaction leads to further fragmentation, which can only be observed in larger system sizes. Finally, we predict a sudden increase in the fragmentation behavior for larger pump intensities.
    \end{abstract}

    \maketitle


    \section{INTRODUCTION}

\begin{figure}[ht!]
    \centering
    \includegraphics[width=\columnwidth]{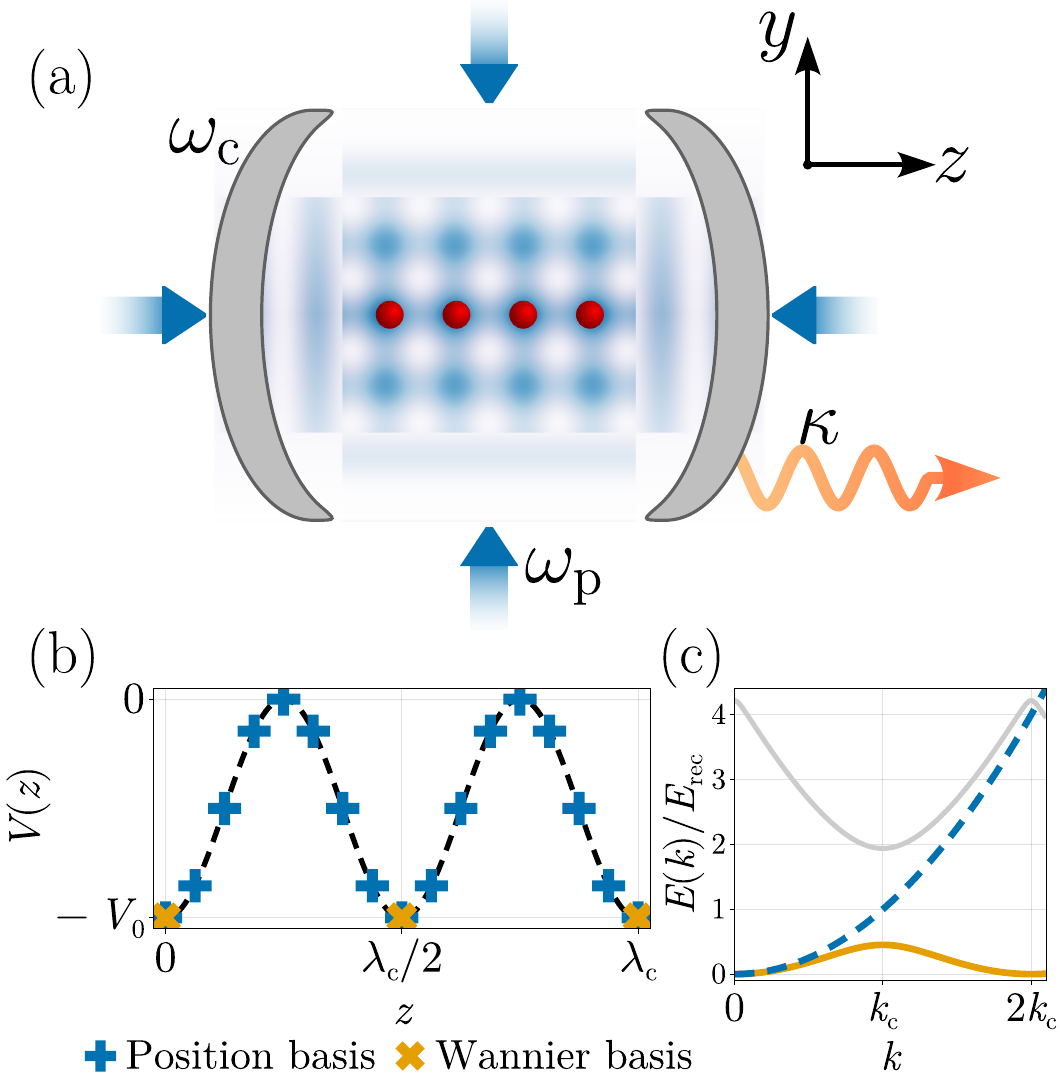}
    \caption{(a) Schematic of the atom-cavity setup with an optical lattice. Ultracold atoms are placed inside a high-finesse optical cavity operating in the good-cavity regime, and are driven using a pump field on the $y$-$z$ plane. The cavity axis is aligned along the $z$-axis. (b) Illustration of the spatial discretization according to the position and (single-band) Wannier basis emphasizing the cavity potential $V(z)$ experienced by a boson. (c) The energies resolved by the (dashed) position and (solid) Wannier bases with $k_\text{c}$ the cavity-mode wave vector. The gray solid curve corresponds to the second band neglected in the standard single-band Wannier expansion.}
    \label{fig:setup}
\end{figure}

Exploring novel phenomena involving quantum many-body systems can be done by studying highly controllable, simplified models realized in cold atom systems \cite{ritsch-2013}. In particular, atom-cavity systems, in which cold atoms are placed inside a high-finesse optical cavity, provide a natural test bed for investigating physics arising from the interactions between light and matter \cite{raimond-2001,walther-2006,mivehvar-2021}. For example, the transversely pumped atom-cavity system can emulate the celebrated Dicke model, as the normal-superradiant phase transition is realized in the transition between a homogeneous superfluid (SF) and self-organized density-wave (DW) phase \cite{domokos-2002,black-2003,baumann-2010,baumann-2011,klinder-2015}. In this case, the Bragg scattering of photons from the transverse pump to the cavity leads to a dynamical optical lattice for the atoms \cite{domokos-2002}. An additional laser field along the cavity axis, as depicted in Fig.~\ref{fig:setup}(a), can be applied to create a static optical lattice which allows for independent tunability of the short-range interactions. 

The additional static optical lattice along the cavity motivates the use of a tight-binding approximation and basis set expansion using Wannier functions, typically restricted to the lowest band of the energy dispersion \cite{jaksch-1998}. This yields a Dicke-Bose-Hubbard model, and its single-band limit has been used to predict interesting phenomena and many-body phases \cite{bakhtiari-2015,landig-2016,dogra-2016,sundar-2016,flottat-2017,himbert-2019,chanda-2021,sharma-2022,orso-2025,halati-2025}. Furthermore, this has been used to study atom-cavity systems with optical lattices having wavelengths incommensurate with the cavity-mode wavelength \cite{habiban-2013}, dynamical gauge potentials using spinor bosons \cite{collela-2022}, and to predict many-body localization in one-dimensional bosonic systems \cite{sierant-2019}. However, certain effects necessitate theories that go beyond the single-band approximation. For example, 
a second energy band is crucial in $p$-band superfluidity in cold atoms in optical lattices \cite{wirth-2011,wang-2021}, and condensation in a dark state in the shaken atom-cavity system \cite{skulte-2023}. In electronic systems inside a cavity, enhancement of cavity-mediated interactions has been predicted due to interband coupling \cite{wang-2024}. Matter-wave superradiance \cite{baumann-2010,klinder-2015} leads to excitations of higher momentum modes outside the first Brillouin zone, limiting the applicability of the single-band theory in this case. Aside from explicitly adding a second band in the Hubbard model, higher bands can be included in the theory by expanding the field operators using a plane-wave basis \cite{kessler-2014,klinder-2016,kessler-2019,kongkhambut-2022,gao-2023,nie-2023,tuquero-2024,halati-2025}, which is equivalent to an expansion using the position basis \cite{tuquero-2024}.

In this work, we compare the dynamics of bosons in an optical lattice placed inside a high-finesse cavity obtained by representing the atomic sector using either a single-band Wannier or a position basis, as depicted in Figs.~\ref{fig:setup}(b) and \ref{fig:setup}(c). Expanding the atomic field operators using the single-band Wannier basis has the numerical advantage that it can capture larger system sizes with multiple unit cells. However, the restriction to the lowest energy band means that certain momentum modes are degenerate, i.e., physics involving higher energy bands will not be captured. In contrast, the use of a position basis will allow for inclusion of excitations to momentum states in higher bands at the expense of being limited to small spatial extents as fine discretization of space is required to guarantee convergence.

At the mean-field level, we find that both bases recover the SF, DW, limit cycle (LC) and irregular dynamical (ID) phases. However, we demonstrate that the Wannier basis fails to capture the large sea of irregular dynamics observed in the position basis, and predicts a different kind of LC due to the limited number of accessible momentum modes in the single-band approximation. We also employ truncated Wigner approximation (TWA) to numerically simulate the dynamics beyond the mean-field theory. Recently, the TWA has been used with the Wannier basis to investigate the static phases, such as the SF and DW phases, in the good-cavity regime \cite{orso-2025,halati-2025}. Here, focusing on the one-dimensional limit, we compare the results of the TWA for the position and Wannier bases. We find that both bases qualitatively agree on the fragmentation dynamics due to the photon-mediated atomic interactions as quantified by the natural occupations, akin to the fragmented superradiant phase \cite{lode-2017}. We also show that larger system sizes are necessary to describe the fragmentation dynamics of bosons with contact interaction, which generally leads to higher degree of fragmentation. For large pump intensities, we observe a sharp increase in fragmentation, in conjunction with the decrease in standard deviation of the occupation number inferred from the Wigner distribution.

The paper is organized as follows. In Sec.~\ref{sec:theory}, we discuss our system and outline the steps to obtain its semiclassical equations of motion. Then, in Sec.~\ref{sec:mean_field_results}, we investigate the mean-field dynamics of the system and discuss the different phases it can exhibit. In Sec.~\ref{sec:bec_fragmentation}, we use TWA to study the fragmentation dynamics of the Bose-Einstein condensate (BEC) for varying pump intensities. We conclude our work in Sec.~\ref{sec:conclusion}.
    \section{THEORY}\label{sec:theory}

\subsection{System Hamiltonian}

We consider an atom-cavity system consisting of $N$ bosonic atoms with mass $m$ placed inside a high-finesse optical cavity with resonance frequency $\omega_\text{c}$ and photon decay rate of $\kappa$. The cavity operates in the good-cavity regime $\kappa\sim\omega_\text{rec}$, where $\omega_\text{rec}=\hbar k_\text{c}^2/2m$ is the recoil frequency, with a cavity mode having a characteristic wavelength $\lambda_\text{c}$ and wave number $k_\text{c}=2\pi/\lambda_\text{c}$. The atoms are driven along the $y$ and $z$ axes using two identical red-detuned pump lasers with frequency $\omega_\text{p}$ and wavelength $\lambda_\text{p}\approx\lambda_\text{c}$. This forms a two-dimensional static optical lattice inside the cavity, the depth of which is controlled by the pump intensity.

We consider the one-dimensional limit, in which the dynamics of the atoms are restricted to the cavity direction. This can be realized by adding a strong confining potential along the direction perpendicular to the cavity axis \cite{maschler-2007}. Then, the effective second-quantized Hamiltonian is \cite{mivehvar-2021,tuquero-2024}
\begin{eqnarray}
    \nonumber \hat{H}(z)&&=-\hbar\Delta_\text{c}\hat{a}^\dagger\hat{a}+\int \hat{\Psi}^\dagger(z)\hat{H}_1(z)\hat{\Psi}(z)\,\dd z \\
    \label{eq:sec02-01} &&+\frac{U_\text{a}}{2}\int\hat{\Psi}^\dagger(z)\hat{\Psi}^\dagger(z)\hat{\Psi}(z)\hat{\Psi}(z)\,\dd z.
\end{eqnarray}
Here, $\hat{a}$ ($\hat{a}^\dagger$) is the intracavity photon annihilation (creation) operator, $\hat{\Psi}(z)$ [$\hat{\Psi}^\dagger(z)$] is the bosonic annihilation (creation) field operator, $\Delta_\text{c}=\omega_\text{p}-\omega_\text{c}$ is the pump-cavity detuning with $\Delta_\text{c}<0$, and $U_\text{a}$ is the atom-atom interaction strength. Lastly, $\hat{H}_1(z)$ is the single-particle effective Hamiltonian given by
\begin{eqnarray}
    \nonumber \hat{H}_1(z)=&&-\frac{\hbar^2}{2m}\,\pdv[2]{}{z}+V(z)+\hbar U_0\hat{a}^\dagger\hat{a}\cos^2(k_\text{c}z) \\%
    \label{eq:sec02-02} &&+\sqrt{\hbar\abs{U_0V_0}}(\hat{a}+\hat{a}^\dagger)\cos(k_\text{c}z),
\end{eqnarray}
where $V(z)=-V_0\cos^2(k_\text{c}z)$ is the potential of the static lattice along the cavity axis, and $U_0$ is the lattice depth of the dynamical lattice for a single photon.

The first term in Eq.~\eqref{eq:sec02-01} denotes the photon sector Hamiltonian, while the last term represents the atom-atom interaction Hamiltonian. The second term in Eq.~\eqref{eq:sec02-01} represents the light-matter Hamiltonian. As shown in Eq.~\eqref{eq:sec02-02}, it contains both the static optical lattice as the second term, and the dynamically generated one in the third term. The last term in Eq.~\eqref{eq:sec02-02} captures the contribution from the interference between the static pump and dynamical cavity fields. We point out that in this 1D limit, the SF phase is expected to only have quasi-long-range order in the presence of contact interaction according to the Mermin-Wagner theorem \cite{mermin-1966,hohenberg-1967}. This is especially relevant later when we consider $U_\text{a}>0$.

\subsection{Atomic field operator expansion}\label{sec:2B}

\begin{figure}[ht!]
    \centering
    \includegraphics[width=\columnwidth]{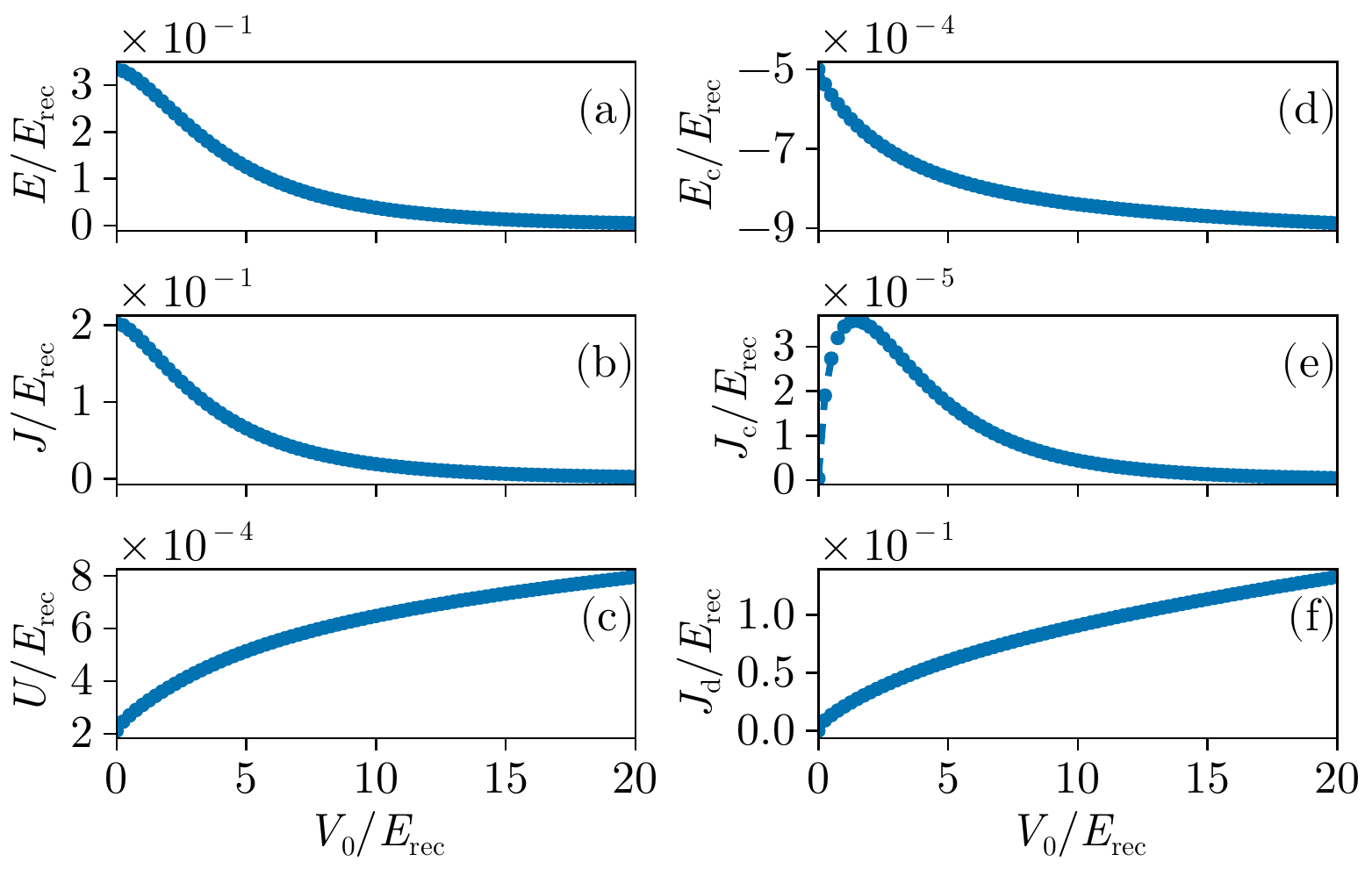}
    \caption{Hubbard parameters for the Wannier basis as a function of $V_0$, obtained from numerically obtained Wannier functions. These were obtained using $U_\text{a}/E_\text{rec}=10^{-3}$ and $U_0/\omega_\text{rec}=-10^{-3}$.}
    \label{fig:wannier_parameters}
\end{figure}

The static optical lattice motivates the use of a tight-binding approximation \cite{jaksch-1998}, which amounts to an expansion of the atomic field operators using Wannier functions. Here, we will restrict the following calculations to Wannier functions $w(z)$ in the lowest energy band of the static lattice $V(z)$. Using the \textit{(single-band) Wannier basis}, the field operators are expanded as
\begin{equation}
    \label{eq:sec02-03} \hat{\Psi}(z)=\sum_jw(z-z_j)\hat{b}_j, \quad%
    \hat{\Psi}^\dagger(z)=\sum_jw(z-z_j)\hat{b}^\dagger_j,
\end{equation}
where $z_j=j\lambda_\text{c}/2$ is the position of the $j$th site and $\hat{b}_j$ ($\hat{b}^\dagger_j$) is the bosonic annihilation (creation) operator for site $j$. Substituting Eq.~\eqref{eq:sec02-03} into Eq.~\eqref{eq:sec02-01}, assuming only nearest-neighbor tunneling, and applying periodic boundary conditions, we obtain a tight-binding Hamiltonian that reads as
\begin{eqnarray}
    \nonumber \hat{H}=&&-\hbar\Delta_\text{c}\hat{a}^\dagger\hat{a}-(J+J_\text{c}\hat{a}^\dagger\hat{a})\sum_j(\hat{b}^\dagger_{j+1}\hat{b}_j+\text{H.c.}) \\%
    \nonumber &&+\frac{U}{2}\sum_j\hat{n}_j(\hat{n}_j-1)+(E+E_\text{c}\hat{a}^\dagger\hat{a})\sum_j\hat{n}_j \\
    \label{eq:sec02-04} &&+J_\text{d}(\hat{a}+\hat{a}^\dagger)\sum_j(-1)^j\hat{n}_j.
\end{eqnarray}
Here, $\hat{n}_j=\hat{b}^\dagger_j\hat{b}_j$ is the on-site number operator for the $j$th lattice site, $E$ is the on-site single-particle energy, $J$ is the tunneling strength, and $U$ is the on-site interaction strength \cite{jaksch-1998}. The dynamical optical lattice leads to cavity-induced shifts to the usual Bose-Hubbard parameters, namely the on-site energy shift $E_\text{c}$ and tunneling energy $J_\text{c}$ \cite{bakhtiari-2015}. The light-matter coupling strength responsible for self-organization into the superradiant DW phase is $J_\text{d}$. In Fig.~\ref{fig:wannier_parameters}, we present the Bose-Hubbard parameters obtained from substituting Eq.~\eqref{eq:sec02-03} into Eq.~\eqref{eq:sec02-01} (see Appendix \ref{sec:wannier_parameters} for more details). Increasing the pump intensity $V_0$ suppresses tunneling between neighboring sites $J$ but enhances the light-matter coupling strength $J_\text{d}$. In Ref.~\cite{landig-2016}, terms proportional to $E_\text{c}$ and $J_\text{c}$ are neglected, which is justified by their comparatively small magnitudes, as illustrated in Fig.~\ref{fig:wannier_parameters}. In our work, we include them for completeness.

Alternatively, the field operators can be expanded in the position (or equivalently, plane-wave) basis as previously done in atom-cavity systems without an additional static cavity lattice \cite{zhiqiang-2017,kessler-2014,klinder-2016,kessler-2019,kongkhambut-2022,gao-2023,nie-2023,tuquero-2024}. This provides a benchmark for dynamics that excite higher energy bands and momentum modes outside the first Brillouin zone, which is the main restriction in the single-band Wannier basis. To this end, in the \textit{position basis}, the field operators are expanded as
\begin{equation}
    \label{eq:sec02-05} \hat{\Psi}(z)=\sum_j\varphi(z-z_j)\hat{c}_j, \quad%
    \hat{\Psi}^\dagger(z)=\sum_j\varphi^*(z-z_j)\hat{c}^\dagger_j,
\end{equation}
where $\hat{c}_j$ ($\hat{c}^\dagger_j$) is the bosonic annihilation (creation) operator for a discretized point in space, and $\varphi(z-z_j)$ is a highly localized function about $z_j$ with
\begin{equation}
    \int\varphi^*(z-z_i)\varphi(z-z_j)\,\dd z=\delta_{i,j},
\end{equation}
and $\partial^2\varphi(z_j)/\partial z^2$ could be approximated using a three-point stencil:
\begin{equation}
    \pdv[2]{\varphi(z_j)}{z}=\frac{\varphi(z_{j+1})-2\varphi(z_j)+\varphi(z_{j-1})}{(\Delta z)^2},
\end{equation}
where $\Delta z$ is the discretization length. The localized state $\varphi(z_j)$ approximates the actual position eigenstate $\delta(z-z_j)$ meaning $\int V(z)\abs{\varphi(z-z_j)}^2\,\dd z\approx V(z_j)$.

For calculations using the position basis, we will limit the simulation box size to only one cavity wavelength $0\leq z_j<\lambda_\text{c}$ which is justified by the periodic boundaries and the absence of a nonperiodic potential. For the rest of this work, we impose the same number of atoms within one cavity wavelength for both Wannier and position bases. Substituting Eq.~\eqref{eq:sec02-05} into Eq.~\eqref{eq:sec02-01}, we obtain a Hamiltonian that reads as
\begin{eqnarray}
    \nonumber \hat{H}=&&-\hbar\Delta_\text{c}\hat{a}^\dagger\hat{a}-J'\sum_j(\hat{c}^\dagger_{j+1}\hat{c}_j+\text{H.c.}) \\%
    \nonumber &&+\frac{U_\text{a}}{2}\sum_j\hat{m}_j(\hat{m}_j-1) \\%
    \nonumber &&+(-V_0+\hbar U_0\hat{a}^\dagger\hat{a})\sum_j\hat{m}_j\cos^2(k_\text{c}z_j) \\%
    \label{eq:sec02-06} && +\sqrt{\hbar\abs{U_0V_0}}(\hat{a}+\hat{a}^\dagger)\sum_j\hat{m}_j\cos(k_\text{c}z_j),
\end{eqnarray}
where $J'=\hbar^2/(2m(\Delta z)^2)$ is the tunneling strength and $\hat{m}_j=\hat{c}^\dagger_j\hat{c}_j$ is the number operator at point $z_j$. 

The single-band Wannier basis is expected to work best if the atoms remain in the first energy band. However, as mentioned earlier, there are dynamical phenomena brought about by nontrivial excitations to higher momentum states at higher energy bands. For example, the $\ket{2\hbar k_\text{c}}$ momentum mode plays an important mode in the activation of the LC phase \cite{kessler-2014,klinder-2016,kessler-2019,kongkhambut-2022,tuquero-2024}. In the single-band Wannier basis, the $\ket{0}$ and $\ket{2\hbar k_\text{c}}$ momentum states are degenerate and, in fact, they correspond to the same state. This already highlights a potential limitation of the single-band Wannier basis for certain phenomena.

\subsection{Semiclassical equations of motion}

The time evolution of a field operator $\hat{O}$ is obtained using the Heisenberg-Langevin equation
\begin{equation}
    \pdv{\expval{\hat{O}}}{t}=\ima\expval{[\hat{H}/\hbar,\hat{O}]+\mathcal{D}\hat{O}},
\end{equation}
where $\hat{H}$ is the system's Hamiltonian, and $\mathcal{D}\hat{O}=\kappa(2\hat{a}^\dagger\hat{O}\hat{a}-\{\hat{a}^\dagger\hat{a},\hat{O}\})$ is the dissipator \cite{breuer}. We are interested in systems with very large particle numbers, which are amenable to mean-field (MF) theory and semiclassical phase-space representations of quantum dynamics \cite{polkovnikov-2010}. Later, we will investigate the degree of fragmentation of the BEC, which requires going beyond standard MF theory. To this end, we utilize the TWA \cite{proukakis,steel-1998}. Within the TWA, we first obtain the Wigner-Weyl transform of Eq.~\eqref{eq:sec02-01}, $H_\text{W}$, using the Bopp representation of the normal ordered operators \cite{polkovnikov-2010}. The dynamics of the system is approximated $c$ fields obtained from the semiclassical equations of motion (EOMs),
\begin{align}
    \label{eq:sec02-08} \ima\hbar\,\pdv{\psi_j}{t}&=\pdv{H_\text{W}}{\psi^*_j}, \\
    \label{eq:sec02-09} \ima\hbar\,\pdv{a}{t}&=\pdv{H_\text{W}}{a^*}-\ima\hbar\kappa a+\ima\hbar\xi,
\end{align}
where $\psi_j$ and $a$ represent the $c$ numbers for the atoms and cavity photons, respectively. The stochastic noise associated with the cavity dissipation satisfies $\langle\xi^*(t)\xi(t')\rangle=\kappa\delta(t-t')$. In the Wannier basis, a system with $L$ lattice sites has $L+1$ coupled stochastic differential equations given by
\begin{eqnarray}
    \nonumber\ima\hbar\,\pdv{a}{t}=&&-\biggl(\hbar\Delta_\text{c}+\ima\hbar\kappa+J_\text{c}\displaystyle\sum_j(b^*_{j+1}b_j+\text{c.c.}) \\
    \nonumber &&-E_\text{c}\displaystyle\sum_j(b^*_jb_j-\tfrac{1}{2})\biggr)a \\
    \label{eq:sec02-10} &&+J_\text{d}\sum_j(-1)^{j-1}(b^*_jb_j-\tfrac{1}{2})+\ima\hbar\xi, \\
    \nonumber\ima\hbar\,\pdv{b_j}{t}=&&-(J+J_\text{c}(a^*a-\tfrac{1}{2}))(b_{j+1}+b_{j-1}) \\
    \nonumber &&+U(b^*_jb_j-1)b_j+(E+E_\text{c}(a^*a-\tfrac{1}{2}))b_j \\
    \label{eq:sec02-11} &&+(-1)^{j-1}J_\text{d}(a+a^*)b_j,
\end{eqnarray}
where $1\leq j\leq L$, while for the position basis with $M$ grid points, the corresponding equations are
\begin{eqnarray}
    \nonumber\ima\hbar\,\pdv{a}{t}=&&-\hbar(\Delta_\text{c}+\ima\kappa)a+\hbar U_0a\sum_j(c^*_jc_j-\tfrac{1}{2})\cos^2(k_\text{c}z_j) \\
    \label{eq:sec02-12} &&+\sqrt{\hbar\abs{U_0V_0}}\sum_j(c^*_jc_j-\tfrac{1}{2})\cos(k_\text{c}z_j)+\ima\hbar\xi, \\
    \nonumber\ima\hbar\,\pdv{c_j}{t}=&&-J'(c_{j+1}+c_{j-1})+U_\text{a}(c^*_jc_j-1)c_j \\
    \nonumber&&+[-V_0+\hbar U_0(a^*a-\tfrac{1}{2})]c_j\cos^2(k_\text{c}z_j) \\
    \label{eq:sec02-13} &&+\sqrt{\hbar\abs{U_0V_0}}(a+a^*)c_j\cos(k_\text{c}z_j),
\end{eqnarray}
with $1\leq j\leq M$.

In the TWA, the quantum dynamics of the system is obtained by including quantum noise in the initial state and fluctuations in dissipation of the cavity-photons \cite{orso-2025,tuquero-2022,tuquero-2024}. An ensemble of initial states, sampled from the initial Wigner distribution, are then propagated according to the corresponding EOM \cite{polkovnikov-2010}. The ensemble average is then taken to determine the dynamics of the cavity-photon occupation number $\abs{a}^2=\expval{a^*a}$ and atomic occupation number at given lattice site or position, depending on the chosen basis, $\abs{\psi_j}^2=\expval{\psi^*_j\psi_j}$.

\begin{figure*}[ht!]
    \centering
    \includegraphics[width=\textwidth]{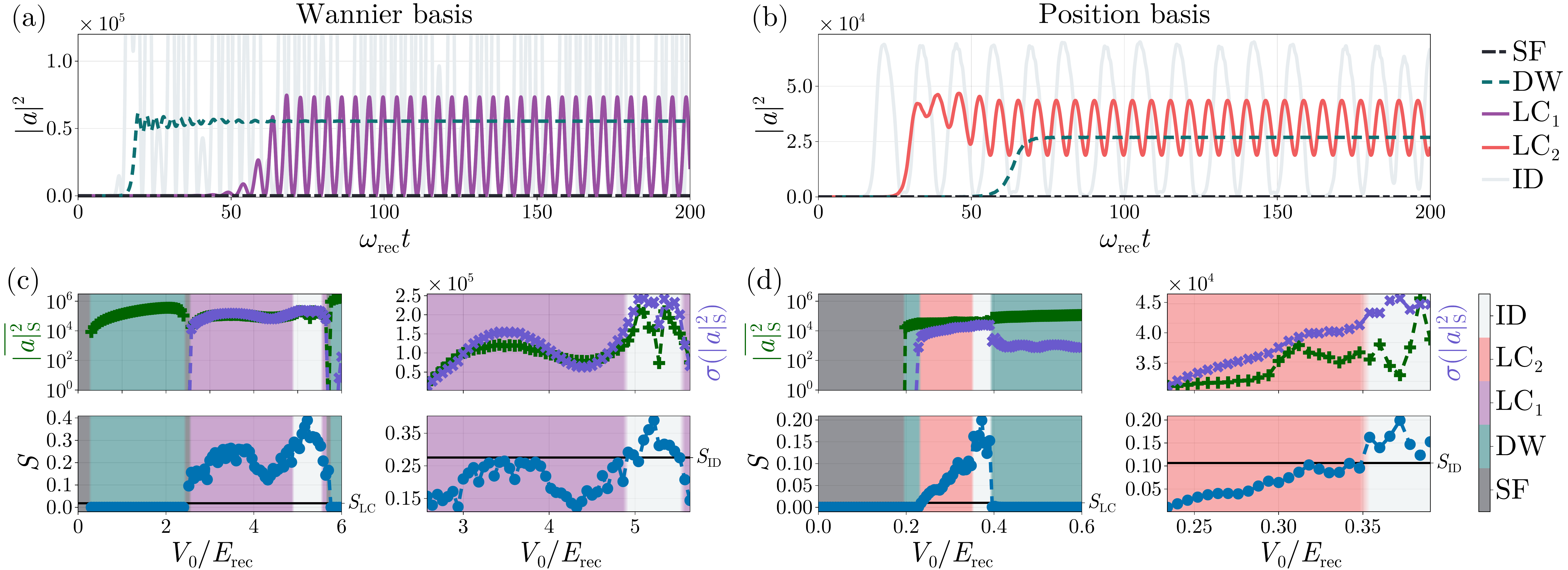}
    \caption{The dynamics of the cavity-photon occupation $\abs{a}^2$ in the SF, DW, LC, and ID phases obtained using the (a) (single-band) Wannier and (b) position basis. The LC phase is labeled as LC$_1$ and LC$_2$ in the Wannier and position basis. The long-time average $\overline{\abs{a}^2_\text{S}}$, standard deviation $\sigma(\abs{a}^2_\text{S})$, and spectral entropy $S$ of $\abs{a}^2_\text{S}$ are used to distinguish between each phase, with (c),(d) showing their behavior in each phase as the pump intensity $V_0$ is increased in the Wannier and position basis, respectively. The black lines in (c),(d) correspond to the threshold spectral entropy values to distinguish the LC, $S_\text{LC}$ and ID, $S_\text{ID}>S_\text{LC}$, phases. Points below the vertical axis in (c),(d) correspond to values close to zero.}
    \label{fig:dynamics}
\end{figure*}

We initially assume an empty cavity containing a uniform cloud of bosonic atoms. The atom cloud is a coherent state that can be sampled numerically using a Gaussian distribution \cite{olsen-2009}, so that for $N$ particles distributed uniformly across $L$ lattice sites and $M$ discretization points in the Wannier and position bases, respectively, under the TWA formalism, the initial cavity-photon field $a(0)$ is
\begin{equation}
    a(0)=\tfrac{1}{2}(\nu_1+\ima\nu_2),
\end{equation}
while the initial atom field $b_j(0)$ and $c_j(0)$ are defined as
\begin{align}
    b_j(0)&=\sqrt{N/L}+\tfrac{1}{2}(\eta_{1,j}+\ima\eta_{2,j}), \\
    c_j(0)&=\sqrt{N/M}+\tfrac{1}{2}(\eta_{1,j}+\ima\eta_{2,j}),
\end{align}
for the Wannier and position bases respectively \cite{olsen-2009}. Here, $\eta_{i,j}$ and $\nu_j$ are random numbers taken from a real normal distribution with $\expval{\eta_{i,j}}=0$, and $\expval{\eta_{i,j}\eta_{i',j'}}=\delta_{i,i'}\delta_{j,j'}$, while $\expval{\nu_j}=0$, and $\expval{\nu_j\nu_{j'}}=\delta_{j,j'}$ \cite{olsen-2009}. Throughout the paper, we consider $N=65\times10^3$ atoms initially distributed evenly across one cavity wavelength, a recoil frequency $\omega_\text{rec}=2\pi\times\qty{3.50}{kHz}$, cavity field decay rate $\kappa=2\pi\times\qty{4.55}{kHz}$, and a dynamical lattice depth $U_0=-2\pi\times\qty{0.360}{Hz}$. For MF simulations we set $U_\text{a}=0$, while later for beyond MF calculations, we will investigate weak short-range interactions $U_\text{a}>0$. For $U_\text{a}=0$ ($U_\text{a}>0$), we use $M=32$ grid points and $L=4$ ($L=32$) lattice sites in the position and Wannier basis respectively (see Appendix \ref{sec:finite_size} for finite-size analysis). The dynamics in the position and Wannier basis are simulated for times $500/\omega_\text{rec}$ and $1500/\omega_\text{rec}$ respectively, with results recorded every $10^{-2}/\omega_\text{rec}$.
    \section{MEAN-FIELD DYNAMICS}\label{sec:mean_field_results}

To illustrate the various phases in the system, we first simulate the MF dynamics for a sudden quench of the pump intensity $V_0$. We obtain the MF dynamics using a single trajectory corresponding to a set of solutions of Eqs.~\eqref{eq:sec02-08} and \eqref{eq:sec02-09} in the absence of stochastic noise $\xi=0$. As the dynamics in this case is deterministic, we numerically integrate the relevant EOM using an adaptive time-step 9(8) Runge-Kutta method \cite{verner-2010}.

\subsection{Static and dynamical phases}

As shown in Figs.~\ref{fig:dynamics}(a) and \ref{fig:dynamics}(b), the system exhibits SF, DW, LC, and ID phases in the MF, each with cavity-photon occupation dynamics consistent with those observed in various experiments \cite{baumann-2011,kessler-2014,klinder-2015-dec,kessler-2019,kongkhambut-2022}. The different phases can be classified based on the long-time dynamics of the cavity-photon occupation $\abs{a}^2$. In the SF phase, the cavity-photon occupation vanishes, while in the DW phase it approaches a constant value. In the LC phase the cavity-photon occupation oscillates at a well-defined frequency, and in the ID phase it exhibits irregular behavior. Based on these defining features, we establish the following order parameters to quantitatively distinguish the different phases in the system: (i) the long-time average of the steady-state cavity-photon occupation $\abs{a}^2_\text{S}$, $\overline{\abs{a}^2_\text{S}}$, (ii) its standard deviation $\sigma(\abs{a}^2_\text{S})$, and (iii) its spectral entropy $S$. These order parameters are numerically obtained from the dynamics in the last $100/\omega_\text{rec}$ of the simulation.

The spectral entropy of the steady-state cavity-photon occupation $\abs{a}^2_\text{S}$ measures the Shannon entropy of its power spectrum $\abs{A}^2_\text{S}(\omega)$ \cite{hosseinzadeh-2007}. The discrete Fourier transform of the steady-state cavity-photon occupation, $\abs{A}^2_\text{S}(\omega)$, is taken. By treating $\abs{A}^2_\text{S}(\omega)$ as a probability distribution, the spectral entropy is calculated using
\begin{equation}
    S=-\sum_i^Np_i\log p_i,
\end{equation}
with $p_i$ given by \cite{hosseinzadeh-2007}
\begin{equation}
    p_i=[\abs{A}^2_\text{S}(\omega_i)]^{-1}\sum_j^N\abs{A}^2_\text{S}(\omega_j).
\end{equation}
The spectral entropy effectively counts the number of peaks present in a signal's power spectrum \cite{misra-2004}.

\begin{figure*}[ht!]
    \centering
    \includegraphics[width=\textwidth]{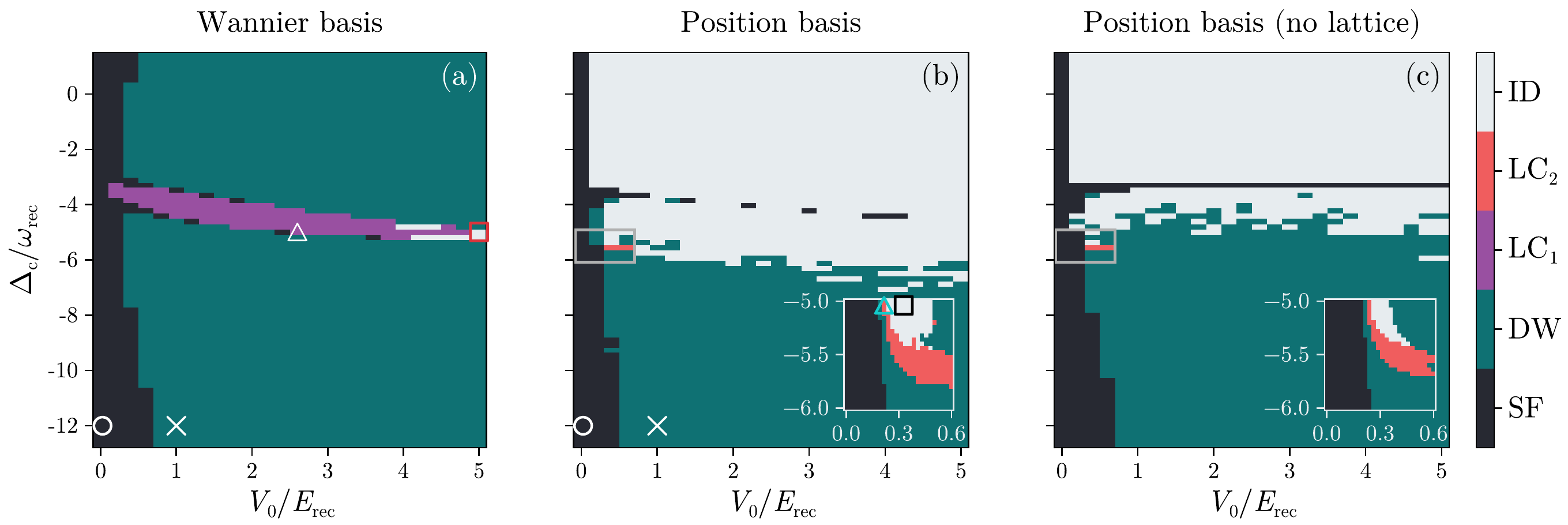}
    \caption{Mean-field phase diagram obtained using the (a)(single-band) Wannier, and position bases in the (b) presence and (c) absence of the static optical lattice. The SF, DW, LC, and ID phases are observed, with LC$_1$ and LC$_2$ corresponding to different LC dynamics in each basis as shown in Fig.~\ref{fig:dynamics}. The insets in (b),(c) show an enlarged view of the boxed region in the main plot. The dotted line marks the matter-wave superradiance regime. The circle, cross, triangle, and square markers mark the values of $\Delta_\text{c}$ and $V_0$ used to obtain Fig.~\ref{fig:no_dynamics}.}
    \label{fig:phase_diagram}
\end{figure*}

We plot the behavior of the order parameters at each phase as the pump intensity is increased in Figs.~\ref{fig:dynamics}(c) and \ref{fig:dynamics}(d). It is seen that in the SF phase, $\overline{\abs{a}^2_\text{S}}\approx0$ and $\sigma(\abs{a}^2_\text{S})\approx0$. In the DW phase for weak pump, the $\overline{\abs{a}^2_\text{S}}$ is a constant value with $\sigma(\abs{a}^2_\text{S})\approx0$, while for strong pump, $\overline{\abs{a}^2_\text{S}}$ is a constant value and $\sigma(\abs{a}^2_\text{S})>0$. Furthermore, we observe that $\overline{\abs{a}^2_\text{S}}$ and $\sigma(\abs{a}^2_\text{S})$ behave similarly in the LC and ID phases.

Ideally, both the standard deviation $\sigma(\abs{a}^2_\text{S})$ and spectral entropy $S$ should vanish in the SF and DW phases since they do not exhibit any oscillations in the long-time limit. However, in the case of the DW phase, we observe transient oscillations that eventually decay but are still included in our calculations of the order parameter (see Appendix \ref{sec:finite_integration}). To account for this, we set a threshold for the maximum value of $\sigma(\abs{a}^2_\text{S})$ and $S$ for a response to be identified as a DW phase. For the LC and ID phases, they exhibit oscillatory dynamics and, in principle, they can be distinguished by the spectral entropy being small for LC and large for ID (see Appendix \ref{sec:decorrelator}). To this end, a response is a DW phase if $\sigma(\abs{a}^2_\text{S})<2.10\times10^4$ (see Appendix \ref{sec:finite_integration}). We also define threshold $S_\text{LC}$ and $S_\text{ID}$ to determine the minimum values of $S$ to enter the LC and ID phases, respectively. In the Wannier basis, we set $S_\text{LC,Wan}=0.0183$ and $S_\text{ID,Wan}=0.275$, while in the position basis, we have $S_\text{LC,pos}=0.0104$ and $S_\text{ID,pos}=0.106$. Different threshold values for the two bases are used due to the distinct cavity-photon dynamics obtained from each basis. Table \ref{tb:orderparameters} summarizes the values of the order parameters corresponding to each dynamical phase of the system, and was used in constructing the MF phase diagrams shown in Fig.~\ref{fig:phase_diagram}.

\begin{table}[ht!]
    \caption{\label{tb:orderparameters} Values of the long-time average, standard deviation, and spectral entropy of the cavity-photon occupation at steady-state, $\overline{\abs{a}^2_\text{S}}$, $\sigma(\abs{a}^2_\text{S})$, and $S$ respectively, for each of the phases exhibited by the system.}
    \begin{ruledtabular}
        \begin{tabular}{l c c c}
            Phase & $\overline{\abs{a}^2_\text{S}}$ & $\sigma(\abs{a}^2_\text{S})$ & $S$ \\ \colrule
            Superfluid (SF)         & 0    & 0    & 0 \\
            Density wave (DW)       & $\geq 10^3$ & $<2.20\times10^4$    & $<S_\text{LC}$ \\
            Limit cycle (LC)        & $>10^3$ & $\geq2.20\times10^4$ & $\geq S_\text{LC}$ \\
            Irregular dynamics (ID) & $>10^3$ & $>2.20\times10^4$ & $\geq S_\text{ID}$
        \end{tabular}
    \end{ruledtabular}
\end{table}

\subsection{Phase diagram}

Calculating the order parameters, as exemplified in Fig.~\ref{fig:dynamics}, allows us to construct the MF phase diagrams shown in Fig.~\ref{fig:phase_diagram} for varying cavity-pump detuning $\Delta_\text{c}/\omega_\text{rec}$ and pump intensity $V_0/E_\text{rec}$ as depicted in Fig.~\ref{fig:phase_diagram}.

The MF phase diagrams using the position basis in Figs.~\ref{fig:phase_diagram}(b) and \ref{fig:phase_diagram}(c) are qualitatively consistent with previous theoretical \cite{tuquero-2024} and experimental \cite{kessler-2019,kongkhambut-2022} results. In contrast, the phase diagram according to the Wannier basis in Fig.~\ref{fig:phase_diagram}(a) displays two qualitatively distinct features compared to those from the position basis, i.e., (i) the appearance of a regular DW phase in regions predicted to be ID by the position basis, and (ii) the type of LC that emerges.

The Wannier basis fails to capture the large sea of irregular dynamics appearing for large $\Delta_\text{c}$. In the position basis, this is demarcated by a horizontal line around $\delta_\text{eff}=\Delta_\text{c}-\frac{1}{2}N\abs{U_0}>0$ as seen in Fig.~\ref{fig:phase_diagram}(c). In the absence of a cavity lattice, this region is identified as matter-wave superradiance, characterized by excitations to higher momentum modes leading to a depletion of the $\ket{0}$ momentum state \cite{baumann-2010}, and is thus a multimode dynamics. Those excitation channels are not included in the single-band Wannier expansion by construction, which thus explains the apparent absence of the large sea of irregular dynamics in the Wannier basis.

\begin{figure}[ht!]
    \centering
    \includegraphics[width=\columnwidth]{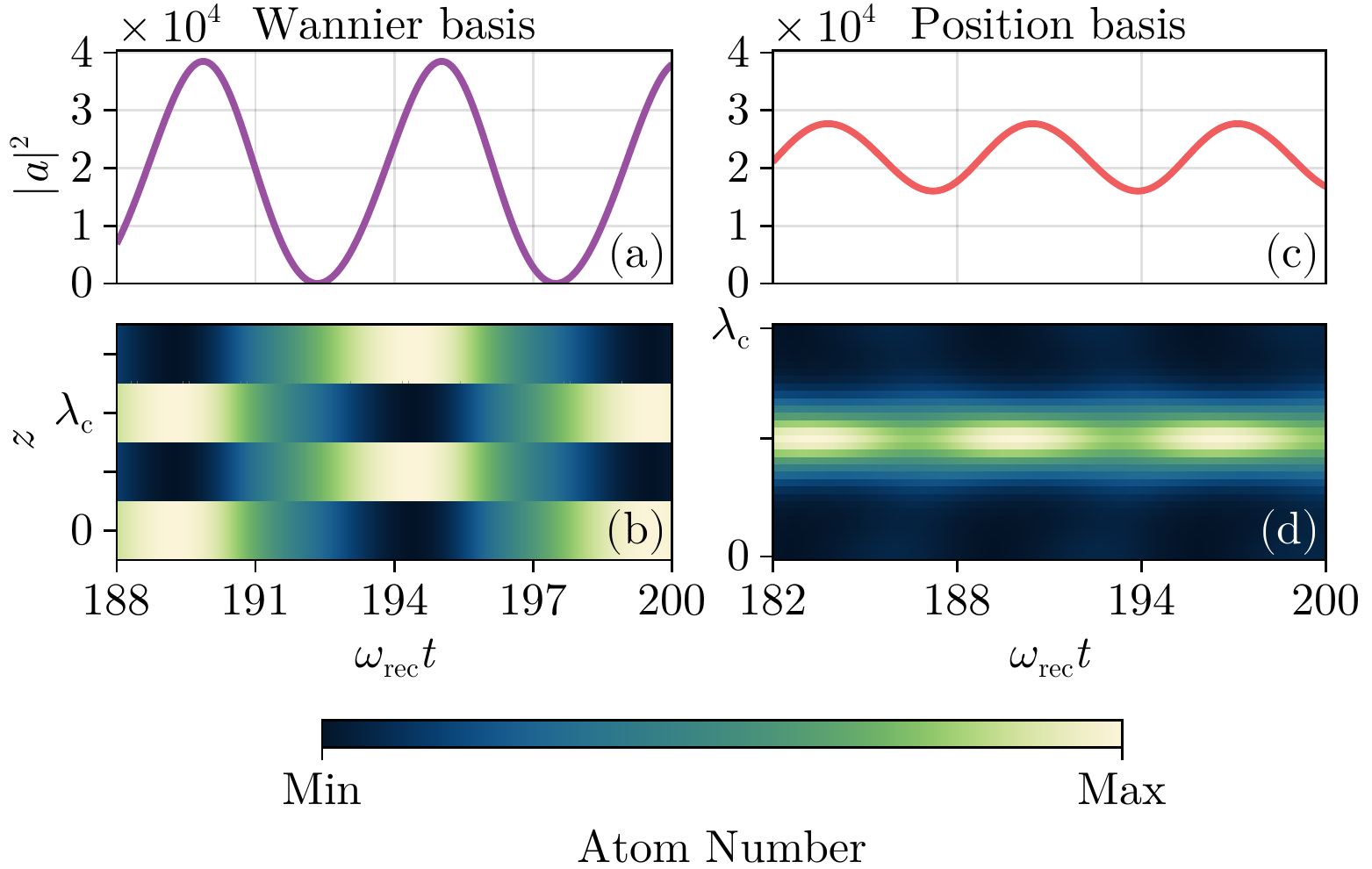}
    \caption{Dynamics of the (a),(c) cavity-photon occupation and (b),(d) single-particle atomic distribution according to the (a),(b) Wannier and (c),(d) position bases. For both bases, $\Delta_\text{c}/\omega_\text{rec}=-5.04$, with $V_0/E_\text{rec}=2.58$ in the Wannier basis, and $V_0/E_\text{rec}=0.216$ in the position basis.}
    \label{fig:lc_dynamics}
\end{figure}

In the LC phase according to the Wannier basis, the cavity-photon occupation oscillates between zero and a large constant. This type of LC is reminiscent of those predicted in Ref.~\cite{kessler-2014} and later observed in Ref.~\cite{gothe-2019}. In Ref.~\cite{kessler-2014}, a two-momentum-mode description was shown to be sufficient for capturing the pulsating dynamics. Note that while the $\ket{2\hbar k_\text{c}}$ momentum mode was included in Ref.~\cite{kessler-2014}, the description is nevertheless two mode only due to the particular choice of probing potential that does not couple to the $\ket{\hbar k_\text{c}}$ momentum mode. In Figs.~\ref{fig:dynamics} and \ref{fig:lc_dynamics}(a), the Wannier basis calculations predict a similar pulsating dynamics of the cavity-photon occupation. This behavior amounts to the system switching between odd and even sites of the lattice as shown in Fig.~\ref{fig:lc_dynamics}(b). That is, the relevant modes according to the single-band Wannier basis for this type of LC are the $\ket{0}$ and $\ket{\pm\hbar k_\text{c}}$ modes. We note that this type of LC is distinct from those observed in Refs.~\cite{kessler-2019,kongkhambut-2022,tuquero-2024,skulte-2024}, which requires the occupation of the second band via the $\ket{\pm2\hbar k_\text{c}}$ modes. It is then tempting to infer from this Wannier basis prediction that the additional static lattice has effectively gapped out the higher energy bands, preventing their occupation by light-matter interactions. However, we point out that higher energy bands are not included in our single-band Wannier expansion by definition, and care is needed before drawing conclusions in this case. In fact, the results of the simulations using the position basis, see Figs.~\ref{fig:dynamics}(b), \ref{fig:lc_dynamics}(c) and \ref{fig:lc_dynamics}(d), recover the type of LC observed in Refs.~\cite{kongkhambut-2022,skulte-2024}, in which the phase of the cavity field is fixed and the atoms are not switching between the odd and even lattice sites. As discussed in Ref.~\cite{skulte-2024}, this type of LC requires at least three sets of momentum modes, $\ket{0}$, $\ket{\pm\hbar k_\text{c}}$, and $\ket{\pm2\hbar k_\text{c}}$ all captured by the position basis, while $\ket{\pm2\hbar k_\text{c}}$ is missed by the single-band Wannier basis by construction.

Comparing the inset plots in Figs.~\ref{fig:phase_diagram}(b) and \ref{fig:phase_diagram}(c), we find that the area in the phase diagram with LC increases with the introduction of the static optical lattice. This can be attributed to how the static lattice enhances the coupling between the $\ket{0}$ and $\ket{\pm2\hbar k_\text{c}}$ modes, with the latter being crucial to the stability of the LC phase. Lastly, we remark that the position and Wannier bases qualitatively agree for more negative $\Delta_\text{c}$, where the dominant transition is SF-DW. In this case, the momentum modes higher than $\ket{\pm\hbar k_\text{c}}$ are not particularly crucial, which suggests the applicability of the single-band Wannier basis in this regime.

We briefly summarize our MF results. Even in the presence of a static optical lattice along the cavity axis, we find that the energy gaps between the bands may not be sufficiently large to prevent excitations to higher momentum modes by light-matter interactions. Concomitantly, this highlights how light-matter interaction can lead to excitations to higher energy bands. This is especially crucial to matter-wave superradiance \cite{baumann-2010} and the type of LC observed in Refs.~\cite{kongkhambut-2022,skulte-2024}. For these phenomena, access to more motional states is required to correctly model them. Within the Wannier expansion, at least one more band is needed in the tight-binding Hamiltonian to account for these effects.
    \section{CONDENSATE FRAGMENTATION}\label{sec:bec_fragmentation}

\begin{figure}[hb!]
    \centering
    \includegraphics[width=\columnwidth]{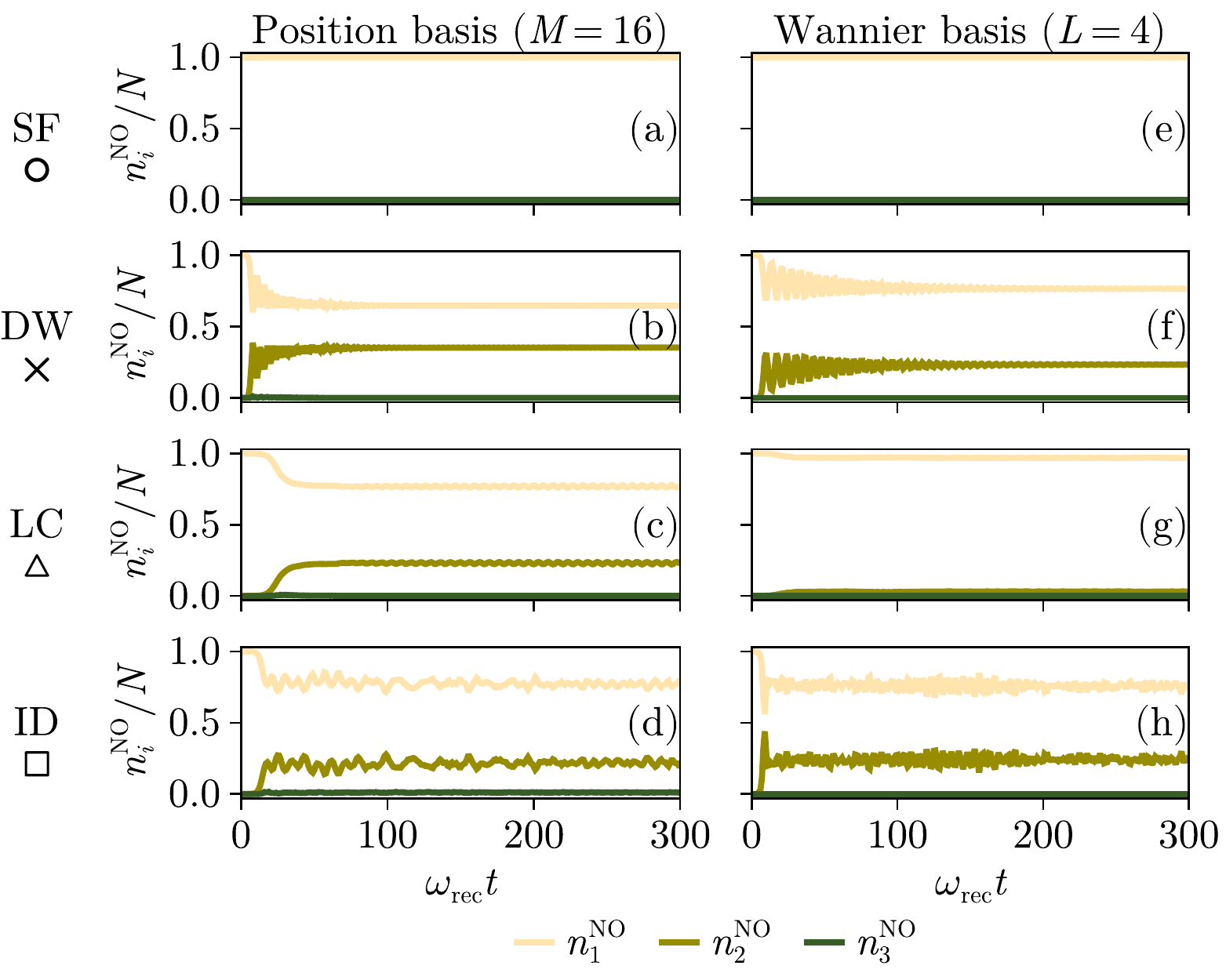}
    \caption{The dynamics of the three largest eigenvalues of the single-particle density matrix with $U_\text{a}=0$ in the SF, DW, LC, and ID phases in the (a)--(d) position and (e)--(h) Wannier basis. The circle, cross, triangle, and square markers correspond to those in Fig.~\ref{fig:phase_diagram}.}
    \label{fig:no_dynamics}
\end{figure}

\begin{figure*}[ht!]
    \centering
    \includegraphics[width=\textwidth]{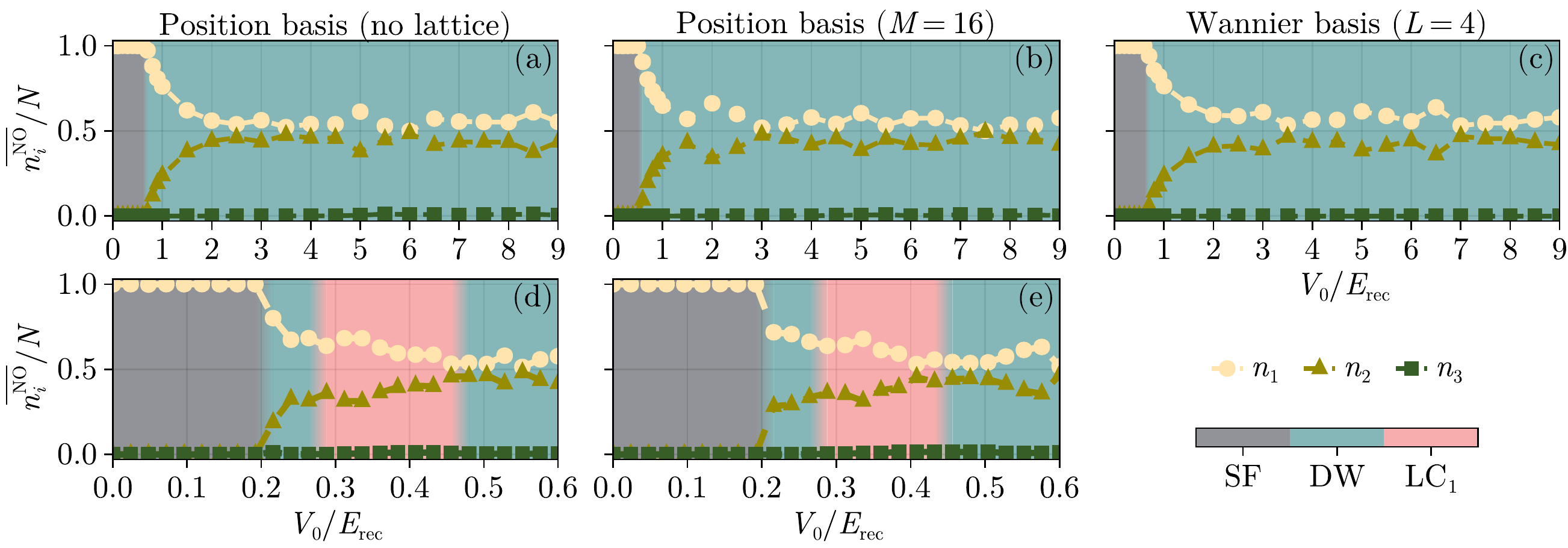}
    \caption{Long-time average of the three largest natural occupations with $U_\text{a}=0$ at steady-state $\overline{n^\text{NO}_i}$ obtained for (a)--(c) SF-DW and (d)--(e) SF-DW-LC phase transitions. In (a)-(c), $\Delta_\text{c}/E_\text{rec}\approx-12.2$; in (d)--(e) $\Delta_\text{c}/E_\text{rec}=-5.4$.}
    \label{fig:no_steady-state}
\end{figure*}

Deeper optical lattices in the atom-cavity system lead to stronger light-matter interactions, which may then increase correlations between the atoms leading to interesting many-body effects. For example, it has been predicted that deeper lattices may fragment the BEC in the DW phase, albeit with atomic contact interactions included in the theory \cite{lode-2017}. Fragmentation of the BEC is inferred from the reduced single-particle density matrix (RSPDM) \cite{penrose-1956} having more than one macroscopic eigenvalue \cite{mueller-2006,bader-2009,lode-2017}.

In the following, we investigate whether condensate fragmentation persists in the absence of inherent short-range interactions meaning the fragmentation is purely driven by long-range photon-mediated interactions. Later, we will also study the effects of including contact interactions on the fragmentation dynamics. To this end, we go beyond mean-field approximation using TWA. Within TWA, we approximate the quantum dynamics by simulating a system of bosonic atoms under a sudden quench in pump intensity $V_0$. We obtain the quantum dynamics by simulating an ensemble of trajectories corresponding to a set of solutions of Eqs.~\eqref{eq:sec02-08} and \eqref{eq:sec02-09} in the presence of stochastic noise $\xi\neq0$, as well as quantum noise in the initial state. We numerically integrate the relevant EOM using a fixed time step Euler-Heun method \cite{iacus} with time step $2^{-14}/\omega_\text{rec}$, and calculate the eigenvalues of the RSPDM.

The RSPDM defined as \cite{mueller-2006}
\begin{equation}
    \label{eq:sec04-01} \rho^{(1)}(z,z')=\expval{\psi^\dagger(z)\psi(z')}
\end{equation}
quantifies long-range order and coherence between the atoms \cite{penrose-1956,nozieres,mueller-2006,bader-2009,lode-2017,lin-2019}. Within TWA, this matrix is constructed from the corresponding elements given by $\rho^{(1)}(z_i,z_j)=\expval{\psi^*(z_i)\psi(z_j)}$ \cite{cosme-2019,skulte-2023}. Its eigenvalues $n^\text{NO}_1\geq n^\text{NO}_2\geq n^\text{NO}_3\geq\dotsm$ and eigenfunctions $\phi^\text{NO}_i(z)$ are known as the natural occupation and natural orbitals respectively \cite{lode-2017}. They measure the fraction of atoms $n^\text{NO}_i$ occupying a quantum state $\phi^\text{NO}_i(z)$ \cite{mueller-2006}. If $\rho^{(1)}(z,z')$  has one macroscopic eigenvalue, the system is a condensate; otherwise it is fragmented \cite{penrose-1956,mueller-2006,bader-2009,lode-2017}. In the following, we limit the ensemble to 50 trajectories (see Appendix \ref{sec:twa_convergence} for details).

\subsection{Photon-induced fragmentation}

In Fig.~\ref{fig:no_dynamics}, we present the dynamics of the three largest natural occupations for representative phases for $U_\text{a}=0$. We find that the natural occupations approach a constant value for long times. In the SF phase $\rho^{(1)}(z,z')$ has one macroscopic eigenvalue. However in the DW, LC, and ID phases, a second natural occupation becomes significant. In those phases, photons are scattered into the cavity, thereby increasing the photon-mediated atom-atom interaction that leads to the BEC further fragmenting. To quantify the fragmentation of the condensate, we calculate the long-time average $\overline{n^\text{NO}_i}$ of the three largest natural occupations within the last $100/\omega_\text{rec}$ of the simulation. Doing this for varying pump intensity yields Fig.~\ref{fig:no_steady-state}. We focus our discussion on the SF-DW and SF-DW-LC phase transitions.

The position and Wannier bases qualitatively agree on the long-time average of the largest natural occupations $\overline{n^\text{NO}_i}$ with increasing pump intensity $V_0/E_\text{rec}$, as seen in Fig.~\ref{fig:no_steady-state}(a)-\ref{fig:no_steady-state}(c). In particular, the degree of fragmentation increases with the pump intensity as seen from the decreasing largest natural occupation $n^\text{NO}_1$, consistent with the results in Ref.~\cite{lode-2017}. However, while Ref.~\cite{lode-2017} includes contact interactions between the bosons, the results in Fig.~\ref{fig:no_steady-state} correspond to bosons with zero contact interaction $U_\text{a}=0$. This suggests that strong photon-mediated atom-atom interactions are enough to make the BEC fragment.

We also observe in Figs.~\ref{fig:no_steady-state}(a), \ref{fig:no_steady-state}(b), \ref{fig:no_steady-state}(d), and \ref{fig:no_steady-state}(e) that the BEC begins to fragment at lower pump intensities in the presence of a static optical lattice. The static optical lattice effectively adds to the potential of the dynamic optical lattice \cite{bakhtiari-2015}. This further suppresses tunneling between neighboring sites compared to the case without the static optical lattice \cite{bakhtiari-2015}; thus destabilizing the SF phase for weaker pump intensities. Moreover, we show in Figs.~\ref{fig:no_steady-state}(d) and \ref{fig:no_steady-state}(e) that the BEC remains fragmented in the LC regime, suggesting a beyond MF character to the previously observed time crystals \cite{kongkhambut-2022,skulte-2024}.

\subsection{Photon- and atomic-induced fragmentation} \label{sec:photon_atom_fragmentation}

Next, we show how the addition of atom-atom interactions and box size affect the fragmentation of the BEC. Throughout this section, we will consider a system of bosons with constant on-site interaction strength, $U_\text{a}/E_\text{rec}=10^{-4}$. In the Wannier basis, this amounts to the on-site interaction varying with the pump strength as depicted in Fig.~\ref{fig:wannier_parameters}(c) [see also Eq.~\eqref{eq:seca01-02}]. In addition to analyzing the natural occupations for varying pump intensity, we also compare the results for $L=4$ and $L=32$ according to the Wannier basis. To this end, we follow the finite-size scaling discussed in Appendix \ref{sec:finite_size}.%

In Fig.~\ref{fig:no_posU}, we show the $\overline{n^\text{NO}_i}$ for varying pump intensity $V_0/E_\text{rec}$ obtained using the position basis and Wannier basis with $L=4$ and $L=32$. We observe that the SF-DW phase transition is pushed to larger pump intensity compared to Fig.~\ref{fig:no_steady-state}, as the bosons need to overcome the repulsive contact interactions to self-organize in the DW phase \cite{landig-2016,dogra-2016,mivehvar-2021}. The position basis predicts that only the first two natural orbitals are significantly occupied. In contrast, in the Wannier basis, more natural orbitals are occupied as the BEC fragments (see Appendix \ref{sec:all_no_with_interaction}).

The occupation of at least one more orbital in Figs.~\ref{fig:no_posU}(b) and \ref{fig:no_posU}(c) suggests that a larger box size is necessary to capture the fragmentation of a system for $U_\text{a}>0$. This highlights a unique advantage of the Wannier basis over the position basis in terms of its scalability to larger system sizes, which as we have shown here is required when short-range atomic interactions play an important role. In Appendix \ref{sec:large_system_no_interaction}, we show that in the absence of atom-atom interactions, the Wannier basis predicts that only two natural orbitals are macroscopically occupied. Thus, Figs.~\ref{fig:no_posU}(b) and \ref{fig:no_posU}(c) elucidate how atom-atom interactions can further increase the correlation between the atoms and that fragmentation increases with the system size. The latter is similar to the decay of long-range order being only resolved if the system size is large enough in 1D systems \cite{mermin-1966,hohenberg-1967}. This also explains how the degree of fragmentation increases with the system size. This is consistent with the picture that the decay in the long-range order is only resolved if the system size is large enough, akin to the quasi-long-range order in 1D systems \cite{mermin-1966,hohenberg-1967}. This explains how the position basis underestimates the fragmentation due to its restriction to a single cavity wavelength.

In Fig.~\ref{fig:no_posU}(c), an apparent transition to a more fragmented state for larger pump intensities can be seen for $V_0/E_\text{rec}>7$. This may hint at an eventual onset of a \textit{superfragmented} DW phase, where more than two natural orbitals are occupied. It is then insightful to analyze the steady-state phase-space distribution of the atoms in the Wannier basis constructed from the distribution of the real and imaginary parts of the corresponding bosonic mode using the solutions of the TWA for long times. For a coherent SF state, the Wigner distribution is Gaussian and thus localized in phase space. On the other hand, an example of an incoherent state, a Fock state, has completely random phase due to the number-phase uncertainty relation and will have ringlike Wigner distribution \cite{cosme-2014}.

\begin{figure}[ht!]
    \centering
    \includegraphics[width=\columnwidth]{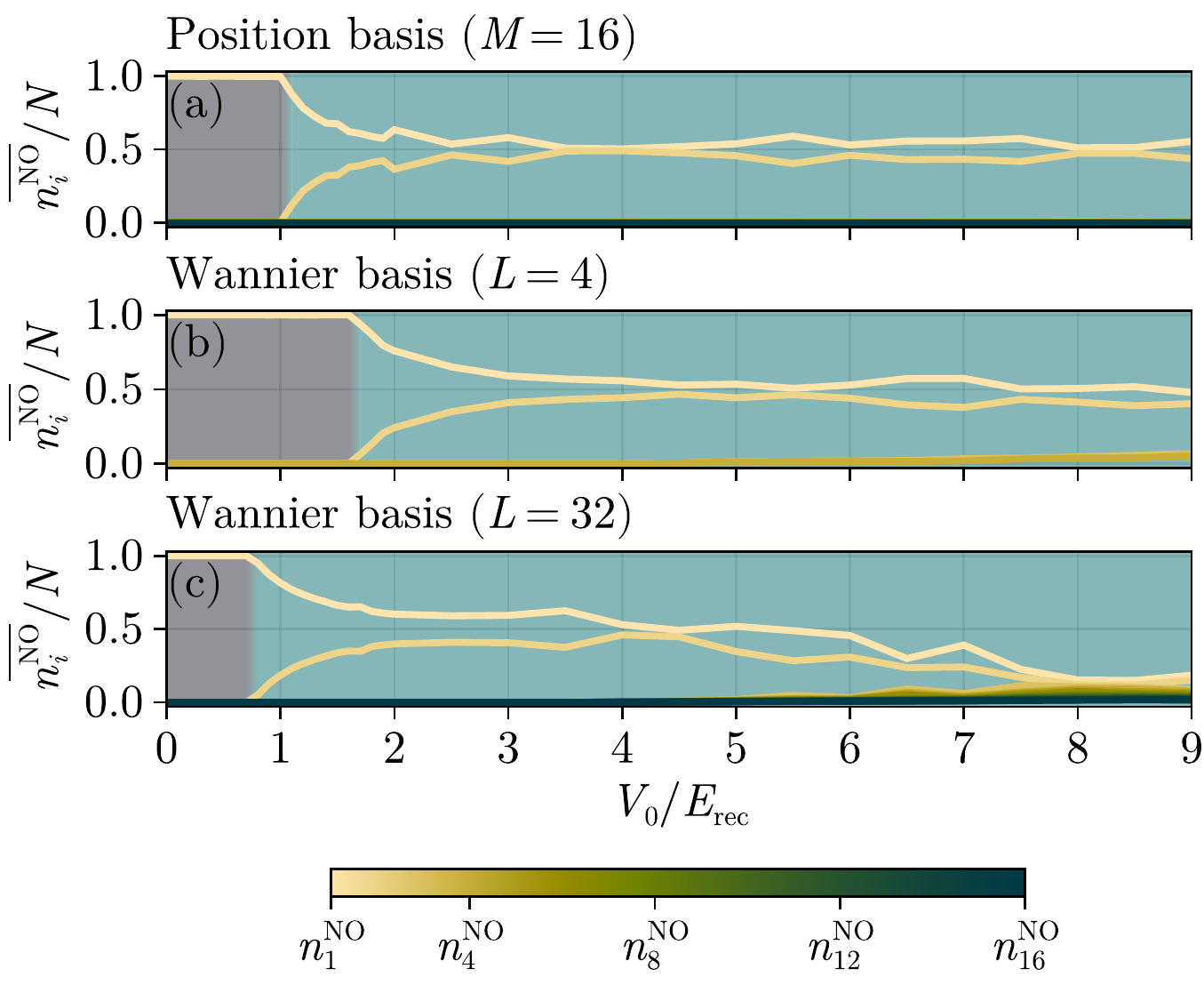}
    \caption{Long-time average of the sixteen largest natural occupations of bosons with contact interaction $U_\text{a}>0$ at steady-state $\overline{n^\text{NO}_i}$ obtained using the (a) position basis, and Wannier basis with $\Delta_\text{c}/E_\text{rec}\approx-12.2$, (b) $L=4$ and (c) $L=32$ lattice sites}
    \label{fig:no_posU}
\end{figure}


In the following, we present the Wigner distributions of one of the symmetry broken states in the DW phase corresponding to the sublattice with a maximum number of atoms in the odd sites, $i=\{1,3,5,\dotsc\}$, rescaled by the maximum magnitude of the cavity occupation number $\abs{b_i}$ in the corresponding trajectory. This is done by only considering trajectories with $\Real(a)<0$  for the first site $i=1$. We also obtain the standard deviation of the occupation number $\sigma_b=\sqrt{(\expval{\hat{n}_1^2}-\expval{\hat{n}_1}^2)/N^2}$, where $N=1.3\times10^5$ for $U_\text{a}=0$ ($L=4$), and $N=10.4\times10^5$ for $U_\text{a}/E_\text{rec}=10^{-4}$ ($L=32$).

\begin{figure}[ht!]
    \centering
    \includegraphics[width=\columnwidth]{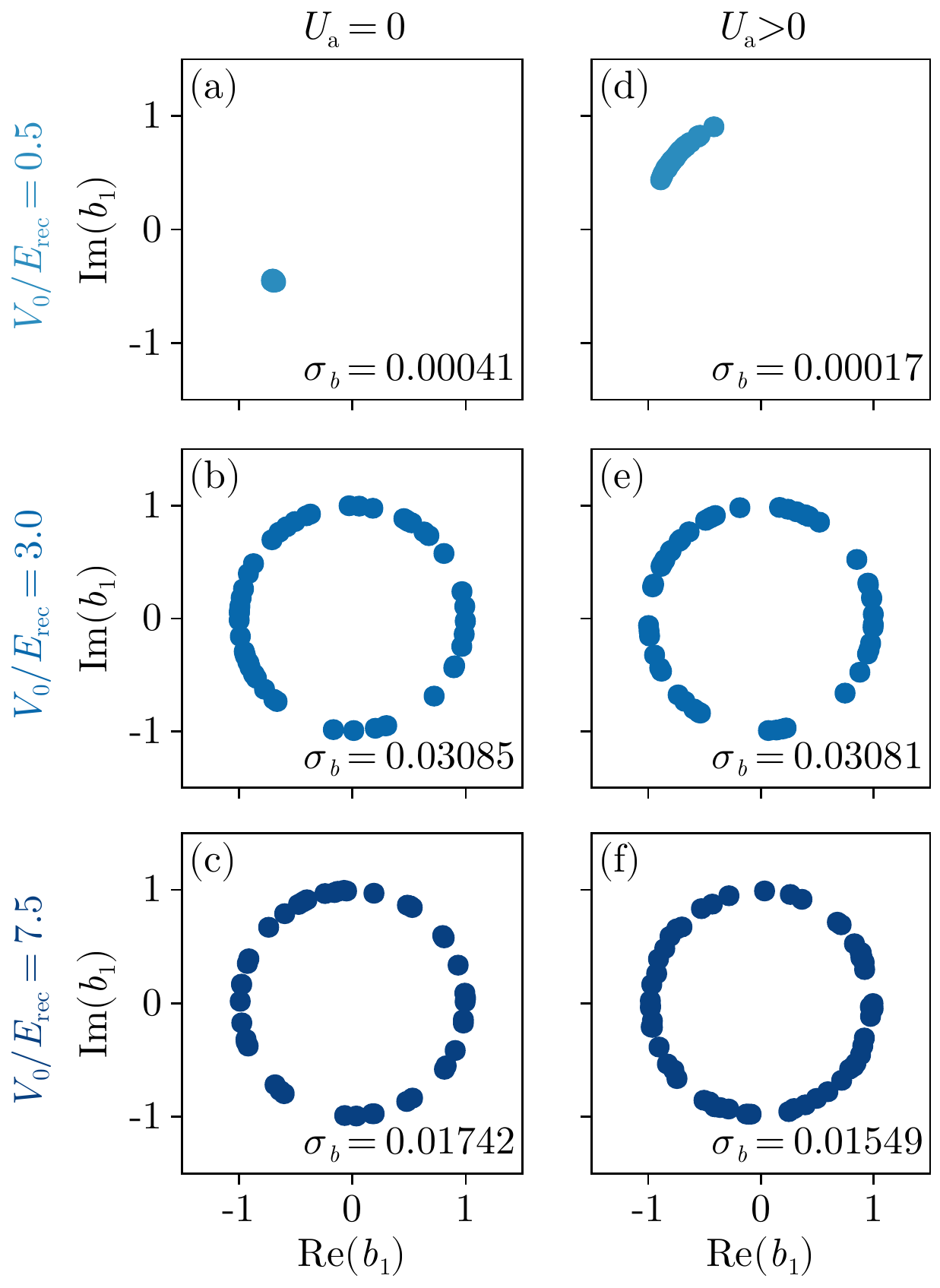}
    \caption{Steady-state Wigner distribution of atoms in the first site for varying pump intensities as indicated. The parameters are $\Delta_\text{c}/E_\text{rec}\approx-12.2$, (a)--(c) $U_\text{a}=0$ and $L=4$, and (d)--(f) $U_\text{a}/E_\text{rec}=10^{-4}$ and $L=32$. The corresponding $\sigma_b$ is indicated.}
    \label{fig:wig_dist}
\end{figure}

In the SF phase, Figs.~\ref{fig:wig_dist}(a) and \ref{fig:wig_dist}(d), we observe that the atomic Wigner distributions are localized in phase space, as expected. A less localized and crescent-shaped distribution is found in Fig.~\ref{fig:wig_dist}(d), which is due to the contact interactions increasing the atomic correlations leading to the Wigner distribution deviating from the ideal Gaussian function.

In Figs.~\ref{fig:wig_dist}(b), \ref{fig:wig_dist}(c), \ref{fig:wig_dist}(e), and \ref{fig:wig_dist}(f), we show the Wigner distributions in the DW phase for $V_0/E_\text{rec}=3$ and $V_0/E_\text{rec}=7.5$. Note that the Wigner distributions in the DW phase exhibit the ringlike distribution reminiscent of a Fock state. However, the distributions have nonzero standard deviation of the occupation number as indicated by $\sigma_b$ in Fig.~\ref{fig:wig_dist}. Nevertheless, as the pump intensity is further increased, $\sigma_b$ decreases. This, together with our observation in Fig.~\ref{fig:no_posU}(c), hints at a possible transition towards a highly-incoherent many-body phase, such as the Mott insulating phase observed in \cite{lin-2021} for stronger pump intensities. This loss of coherence is further corroborated by the increase in phase fluctuations inferred from the distribution of phase differences between adjacent sites, as shown in Appendix \ref{sec:phase_fluctuations}. To further characterize this regime, it is desirable for future works to utilize alternative many-body methods \cite{halati-2020,lin-2020,lode-2017,lin-2021,uwe-2024,molignini-2025}, due to the TWA's limitation in describing strongly correlated systems \cite{proukakis,sinatra-2001,blakie-2008}. Nevertheless, despite the apparent shortcomings of TWA, we point out that TWA and related methods have exhibited signatures of strongly correlated phenomena, such as thermalization and many-body localization \cite{cosme-2014,acevedo-2017}.

Let us summarize our results in this section. In the absence of atom-atom interactions, a strong enough photon-mediated atom-atom interaction can fragment the BEC. This fragmentation is enhanced by the presence of atom-atom interaction. Moreover, we showed that while the position and Wannier basis predict similar fragmentation dynamics for  bosons with $U_\text{a}=0$, the position basis limited to a single cavity wavelength underestimates the degree of fragmentation for $U_\text{a}>0$ in the thermodynamic limit. This is due to the fragmentation increasing with the system size as revealed by the Wannier basis simulations. Finally, for $U_\text{a}>0$, TWA predicts a crossover to a more fragmented state, which may be indicative of an eventual phase transition for large pump intensities.
    \section{CONCLUSION}\label{sec:conclusion}

In conclusion, we have studied the dynamics of bosons inside a high-finesse cavity with laser fields applied along and perpendicular to the cavity axis. Specifically, we compare the dynamics obtained using a single-band Wannier and position basis for the atomic field operators. The single-band Wannier basis is motivated by the presence of an additional static optical lattice and has a computational advantage over the position basis for large system sizes at the expense of only being limited to excitations in the lowest band. In contrast, the position basis includes higher energy bands but is limited to small system sizes due to the necessity of sampling enough discretized points in space. Both bases have recovered the SF, DW, LC, and ID phases reported in the literature \cite{kessler-2014,klinder-2015,klinder-2015-dec,bakhtiari-2015,klinder-2016,kessler-2019,kongkhambut-2022,tuquero-2024}, but the Wannier basis displayed two qualitatively distinct features compared to those from the position basis: (i) the absence of large sea of irregular dynamics, and (ii) the type of LC that emerges, which we attribute to the excitation channels accessible to each basis. In the case of the LC observed in Refs.~\cite{kongkhambut-2022,skulte-2024}, at least three sets of momentum modes, with one of them being in the second band, are necessary \cite{skulte-2024}. Due to the limitation of the single-band Wannier expansion, it is unable to capture these phenomena.

We have also investigated the fragmentation dynamics due to many-body interactions using the TWA. We have analyzed the spectral properties of the RSPDM, with special focus on the natural occupations to quantify the degree of fragmentation. In particular, we have demonstrated that fragmentation persists even in the absence of atomic contact interactions. That is, photon-mediated atom-atom interactions can already lead to correlations that can cause the BEC to fragment. Furthermore, the position and Wannier bases are shown to yield qualitatively similar predictions for the fragmentation of the BEC away from the SF phase. We have demonstrated that larger system sizes are necessary to fully capture the fragmentation dynamics in the presence of contact interaction between bosons. Because of this, the position basis, limited to a single cavity wavelength, underestimates the degree of fragmentation for $U_\text{a}>0$. We have observed using the Wannier basis that contact interaction increases the degree of fragmentation of the BEC, highlighting a computational advantage of the Wannier basis over the position basis due to its scalability for such scenarios. Our TWA simulations in the case when there are contact interactions predict that the system can transition into a more fragmented DW state as the pump intensity is increased.


    \begin{acknowledgments}
        J.G.C acknowledges support from the National Academy of Science and Technology, Philippines - Department of Science and Technology, Philippines.
    \end{acknowledgments}

    \appendix

    \section{OBTAINING THE WANNIER BASIS PARAMETERS}\label{sec:wannier_parameters}

Substituting Eq.~\eqref{eq:sec02-03} into Eq.~\eqref{eq:sec02-01}, assuming only nearest-neighbor tunneling, and applying periodic boundary conditions, the on-site single particle energy $E$, tunneling strength $J$, and on-site interaction strength is
\begin{eqnarray}
    \nonumber E&&=\int w^*(z)\left[ -\frac{\hbar^2}{2m}\pdv[2]{}{z}-V_0\cos^2(k_\text{c}z) \right] \\
    &&\qquad\quad\times w(z)\,\dd z, \\
    \nonumber J&&=-\int w^*(z)\left[ -\frac{\hbar^2}{2m}\pdv[2]{}{z}-V_0\cos^2(k_\text{c}z) \right] \\
    &&\qquad\quad\times w(z-\lambda_\text{c}/2)\,\dd z, \\
    \label{eq:seca01-02} U&&=U_\text{a}\int\abs{w(z)}^4\,\dd z.
\end{eqnarray}
In addition, the cavity-induced on-site energy shift $E_\text{c}$, tunneling energy $J_\text{c}$, and light-matter coupling strength $J_\text{d}$ are
\begin{eqnarray}
    \label{eq:seca01-03} E_\text{c}&&=\hbar U_0\int w^*(z)\cos^2(k_\text{c}z)w(z)\,\dd z, \\
    \label{eq:seca01-04} J_\text{c}&&=-\hbar U_0\int w^*(z)\cos^2(k_\text{c}z)w(z-\lambda_\text{c}/2)\,\dd z, \\
    \label{eq:seca01-05} J_\text{d}&&=\sqrt{\hbar\abs{U_0V_0}}\abs{\int w^*(z)\cos(k_\text{c}z)w(z)\,\dd z}.
\end{eqnarray}
To obtain these parameters, we need to first calculate the Wannier function.

The Wannier function $w(z)$ is obtained from a superposition of Bloch functions corresponding to the static optical lattice potential $V(z)=-V_0\cos^2(k_\text{c}z)$. The Bloch functions for some momentum $k$
\begin{equation}
    \psi_k(z)=\pol{kz}u_k(z),
\end{equation}
satisfy the Schr\"{o}dinger equation:
\begin{equation}
    \label{eq:secA1-07} \left[ -\frac{\hbar^2}{2m}\,\pdv[2]{}{z}-V_0\cos^2(k_\text{c}z) \right]\pol{kz}u_k(z)=E(k)\pol{kz}u_k(z),
\end{equation}
where $E(k)$ is the lowest eigenenergy at momentum $k$. Writing $V(z)=-V_0\cos^2(k_\text{c}z)$ and $u_k(z)$ in terms of its discrete Fourier transform, substituting to Eq.~\eqref{eq:secA1-07}, and simplifying, we get
\begin{eqnarray}
    \nonumber E(k)\varphi_n(k)&&=\frac{\hbar^2}{2m}(k+2nk_\text{c})^2\varphi_n(k) \\
    &&\quad-\tfrac{1}{4}V_0(\varphi_{n-1}(k)+\varphi_{n+1}(k)),
\end{eqnarray}
where $\varphi_n(k)$ is the eigenstate corresponding to the lowest eigenenergy at momentum $k$. Thus, $E(k)$ and $\varphi_n(k)$ are obtained from the eigenvalues and eigenstates of the matrix
\begin{equation}
    \label{eq:secA1-09} \tilde{E}_{m,n}=\frac{\hbar^2}{2m}(k+2nk_\text{c})^2\delta_{m,n}-\tfrac{1}{4}V_0[\delta_{n,n-1}+\delta_{n,n+1}].
\end{equation}
Moreover, because $V(z)$ has a spatial period of $a=2\pi/(2k_\text{c})$, the first Brillouin zone lies within $-k_\text{c}\leq k<k_\text{c}$.

The Wannier state corresponding to the lowest energy $\ket{l}$ is
\begin{equation}
    \ket{l}\equiv\frac{1}{\sqrt{L_\text{sys}}}\sum_k\pol[-]{akl}\ket{k},
\end{equation}
where $L_\text{sys}$ is the system size, and $\ket{k}$ is the single-particle Bloch state corresponding to the lowest energy given by
\begin{equation}
    \ket{k}=\int\sum_n\varphi_n(k)\pol{(k+2nk_\text{c}z)}\ket{z}\,\dd z.
\end{equation}
The Wannier function for the $l$th site is then,
\begin{equation}
    w(z-la)=\braket{z}{l}=\frac{1}{\sqrt{L_\text{sys}}}\sum_k\left[\sum_n\varphi_n(k)\pol{nk_\text{c}z}\right]\pol{k(z-al)}.
\end{equation}
    \section{FINITE-SIZE SCALING ANALYSIS}\label{sec:finite_size}

We use finite-size scaling to guarantee convergence of our numerical simulations for larger system sizes. To this end, we rescale the $c$ numbers for the atoms $\psi_j$ and cavity photons $a$ by the number of atoms $N$ in the system
\begin{equation}
    \label{eq:seca02-01} a\to a/\sqrt{N}, \quad\text{and}\quad \psi_j\to\psi_j/\sqrt{N}.
\end{equation}

Applying Eq.~\eqref{eq:seca02-01} to Eqs.~\eqref{eq:sec02-10} and \eqref{eq:sec02-11}, the rescaled semiclassical equations of motion in the Wannier basis are
\begin{eqnarray}
    \nonumber\ima\hbar\,\pdv{\alpha}{t}=&&-\biggl(\hbar\Delta_\text{c}+\ima\hbar\kappa+J_\text{c}N\displaystyle\sum_j(\beta^*_{j+1}\beta_j+\text{c.c.}) \\
    \nonumber &&-E_\text{c}N\displaystyle\sum_j(\beta^*_j\beta_j-\tfrac{1}{2N})\biggr)\alpha \\
    \label{eq:seca02-02} &&+J_\text{d}\sqrt{N}\sum_j(-1)^{j-1}(\beta^*_j\beta_j-\tfrac{1}{2}), \\
    \nonumber\ima\hbar\,\pdv{\beta_j}{t}=&&-[J+J_\text{c}N(\alpha^*\alpha-\tfrac{1}{2N})](\beta_{j+1}+\beta_{j-1}) \\
    \nonumber&&+UN(\beta^*_j\beta_j-\tfrac{1}{N})\beta_j+[E+E_\text{c}N(\alpha^*\alpha-\tfrac{1}{2N})]\beta_j \\
    \label{eq:seca02-03} &&+(-1)^{j-1}J_\text{d}\sqrt{N}(\alpha+\alpha^*)\beta_j,
\end{eqnarray}
where $1\leq j\leq L$, $\alpha=a/\sqrt{N}$ and $\beta_j=b_j/\sqrt{N}$. As seen in Eqs.~\eqref{eq:seca01-03}--\eqref{eq:seca01-05}, $E_\text{c}$, $J_\text{c}$, and $J_\text{d}$ are all proportional to $U_0$.

Alternatively, applying Eq.~\eqref{eq:seca02-01} to Eqs.~\eqref{eq:sec02-12} and \eqref{eq:sec02-13}, we have for the position basis:
\begin{eqnarray}
    \nonumber\ima\hbar\,\pdv{\alpha}{t}=&&-\hbar(\Delta_\text{c}+\ima\kappa)\alpha \\
    \nonumber &&+\hbar U_0N\alpha\sum_j(\chi^*_j\chi_j-\tfrac{1}{2N})\cos^2(k_\text{c}z_j) \\
    \label{eq:seca02-04} &&+\sqrt{\hbar\abs{U_0NV_0}}\sum_j(\chi^*_j\chi_j-\tfrac{1}{2N})\cos(k_\text{c}z_j), \\
    \nonumber\ima\hbar\,\pdv{\chi_j}{t}=&&-J'(\chi_{j+1}+\chi_{j-1})+UN(\chi^*_j\chi_j-\tfrac{1}{N})\chi_j \\
    \nonumber&&+[-V_0+\hbar U_0N(\alpha^*\alpha-\tfrac{1}{2N})]\chi_j\cos^2(k_\text{c}z_j) \\
    \label{eq:seca02-05} &&+\sqrt{\hbar\abs{U_0NV_0}}(\alpha+\alpha^*)\chi_j\cos(k_\text{c}z_j),
\end{eqnarray}
where $1\leq j \leq M$ and $\chi_j=c_j/\sqrt{N}$.

For finite-size scaling analysis, we simulate the MF dynamics of a sudden quench of the pump intensity $V_0$. We obtain the MF dynamics using a single trajectory corresponding to a set of solutions of Eqs.~\eqref{eq:seca02-02} and \eqref{eq:seca02-03} or Eqs.~\eqref{eq:seca02-04} and \eqref{eq:seca02-05}. We vary the system size, but keep $N/L_\text{sys}$, where $L_\text{sys}$ is the total system size, $NU_0$, and $NU_\text{a}$ fixed. We initially consider $M=32$ grid points and $L=4$ lattice sites in the position and Wannier basis respectively. The same parameters mentioned in Sec.~\ref{sec:theory} were used, with pump strength $V_0/E_\text{rec}=1$, and bare cavity detuning $\Delta_\text{c}/E_\text{rec}\approx-12.2$. We numerically integrate the relevant EOM and compare the rescaled cavity-photon occupation number $\abs{a}^2/N$.

\begin{figure}[!ht]
    \centering
    \includegraphics[width=\columnwidth]{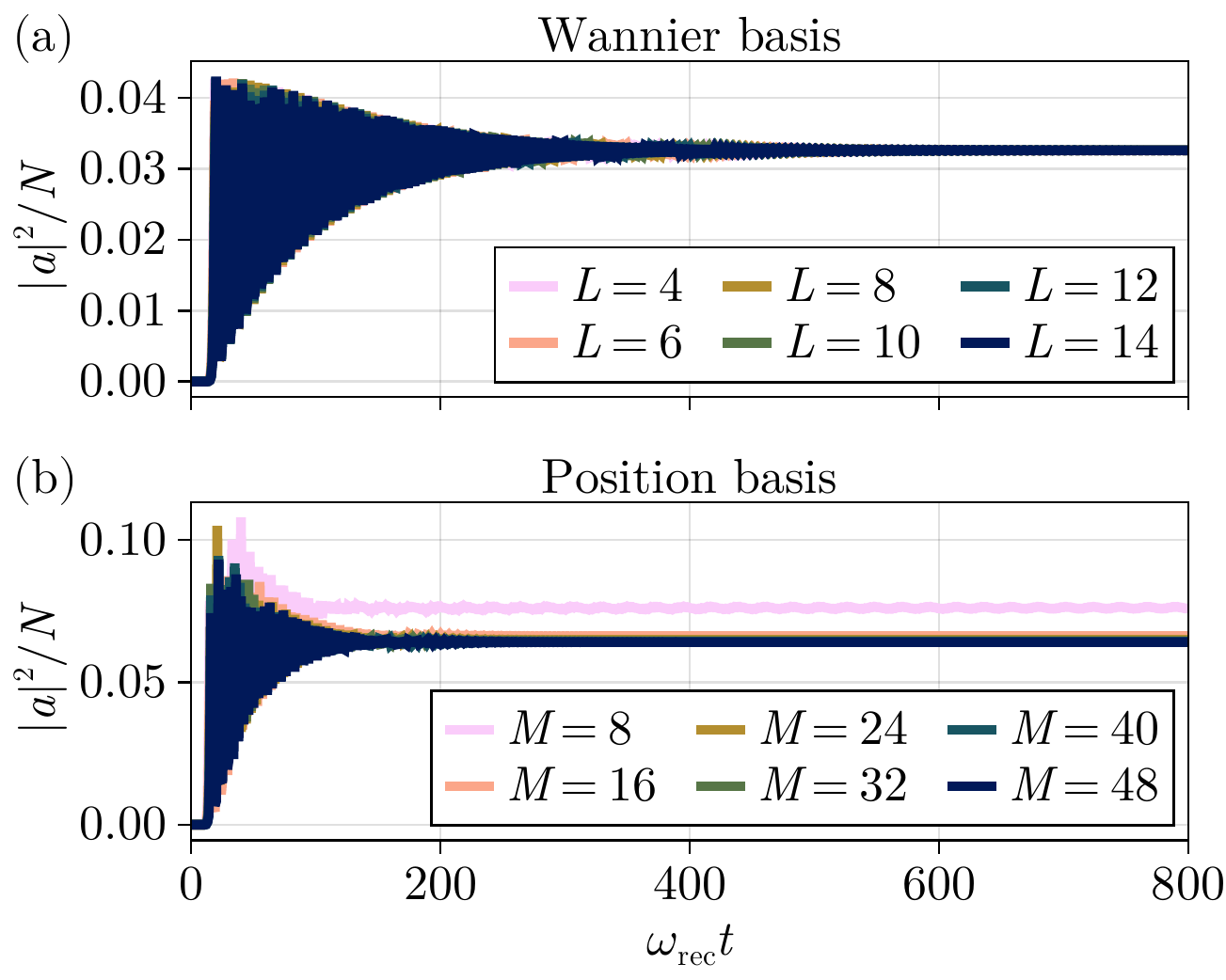}
    \caption{Rescaled cavity-photon occupation number $\abs{a}^2/N$ dynamics for different system sizes for (a) Wannier and (b) position basis with $U_\text{a}=0$.}
    \label{fig:fss_comparison}
\end{figure}

We first consider $U_\text{a}=0$. The relevant parameters are $NU_0=-2\pi\times23.4\times10^3$, with fixed $N/L=32.5\times10^3$ and $N/M=2.03\times10^3$ for the Wannier and position bases, respectively. In Fig.~\ref{fig:fss_comparison}, we show the dynamics of the rescaled cavity-photon occupation number $\abs{a}^2/N$ dynamics for varying system sizes. As expected, the results converge as the system size is increased. The long-time average of $\abs{a}^2/N$ in the Wannier basis is half of that in the position basis because we initially set $L_\text{sys}=2\lambda_\text{c}$ in the Wannier basis, which contains twice the number of atoms across one cavity wavelength. Our results provide us a benchmark of the minimal system size to simulate the dynamics of a large system. As seen in Fig.~\ref{fig:fss_comparison}, at least $L=4$ lattice sites are necessary to guarantee convergence using the Wannier basis, while at least $M=16$ discretization points are needed to capture $L_\text{sys}=\lambda_\text{c}$ using the position basis. This highlights the numerical advantage of the Wannier basis compared to the position basis as it can capture larger system sizes with multiple unit cells.

\begin{figure}[!ht]
    \centering
    \includegraphics[width=\columnwidth]{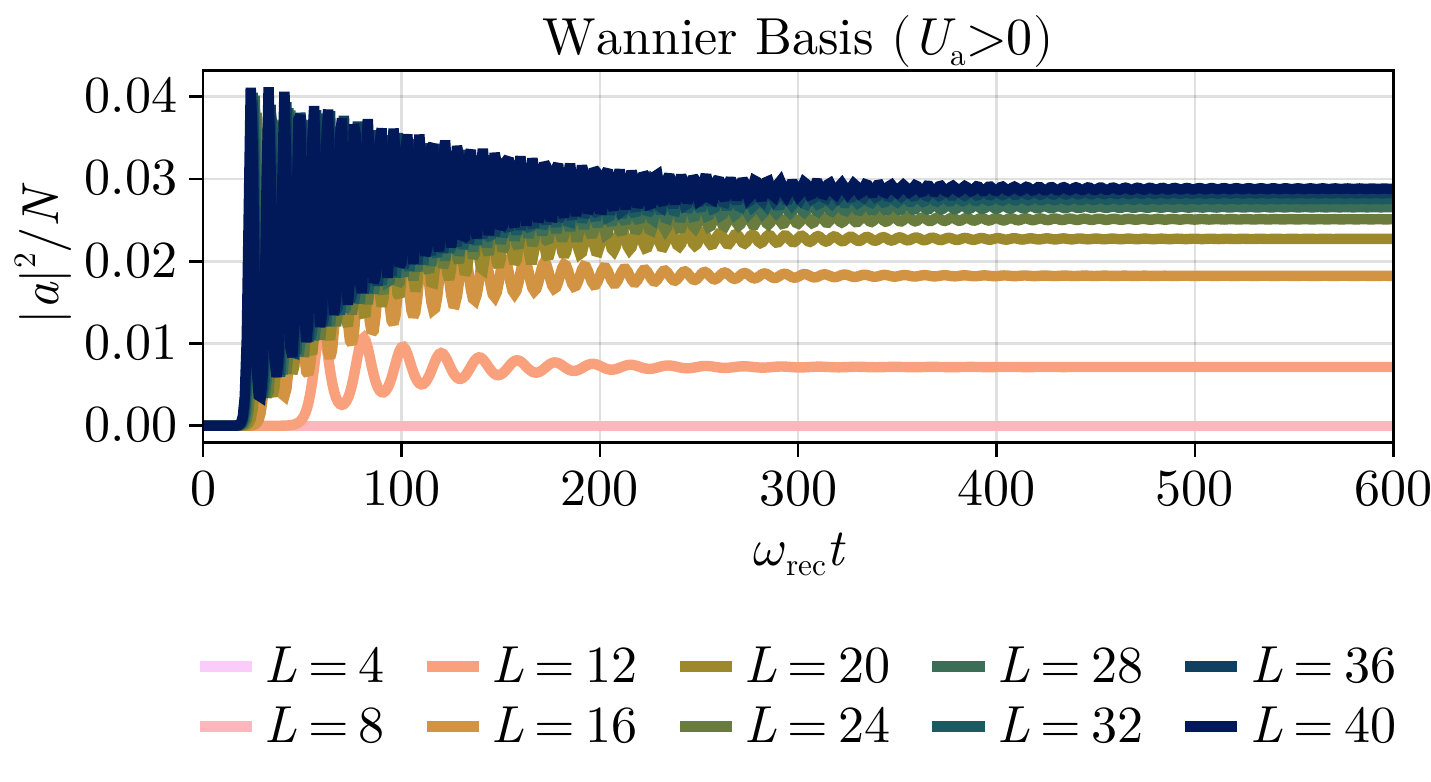}
    \caption{Rescaled cavity-photon occupation number $\abs{a}^2/N$ dynamics for different system sizes for (a) Wannier and (b) position basis with $U_\text{a}/E_\text{rec}=10^{-4}$.}
    \label{fig:fss_comparison_posU}
\end{figure}

We now consider $U_\text{a}/E_\text{rec}=10^{-4}$. In addition to $N/L=32.5\times10^3$ and $NU_0=-2\pi\times23.4\times10^3$, we fix $NU_\text{a}/E_\text{rec}=13$. In Fig.~\ref{fig:fss_comparison_posU}, we show the dynamics of $\abs{a}^2/N$ according to the Wannier basis. We see that at least $L=32$ lattice sites, which correspond to a total system size of $L_\text{sys}=(32/2)\lambda_\text{c}=16\lambda_\text{c}$, are required to guarantee convergence.
    \begin{figure*}[ht!]
    \centering
    \includegraphics[width=\textwidth]{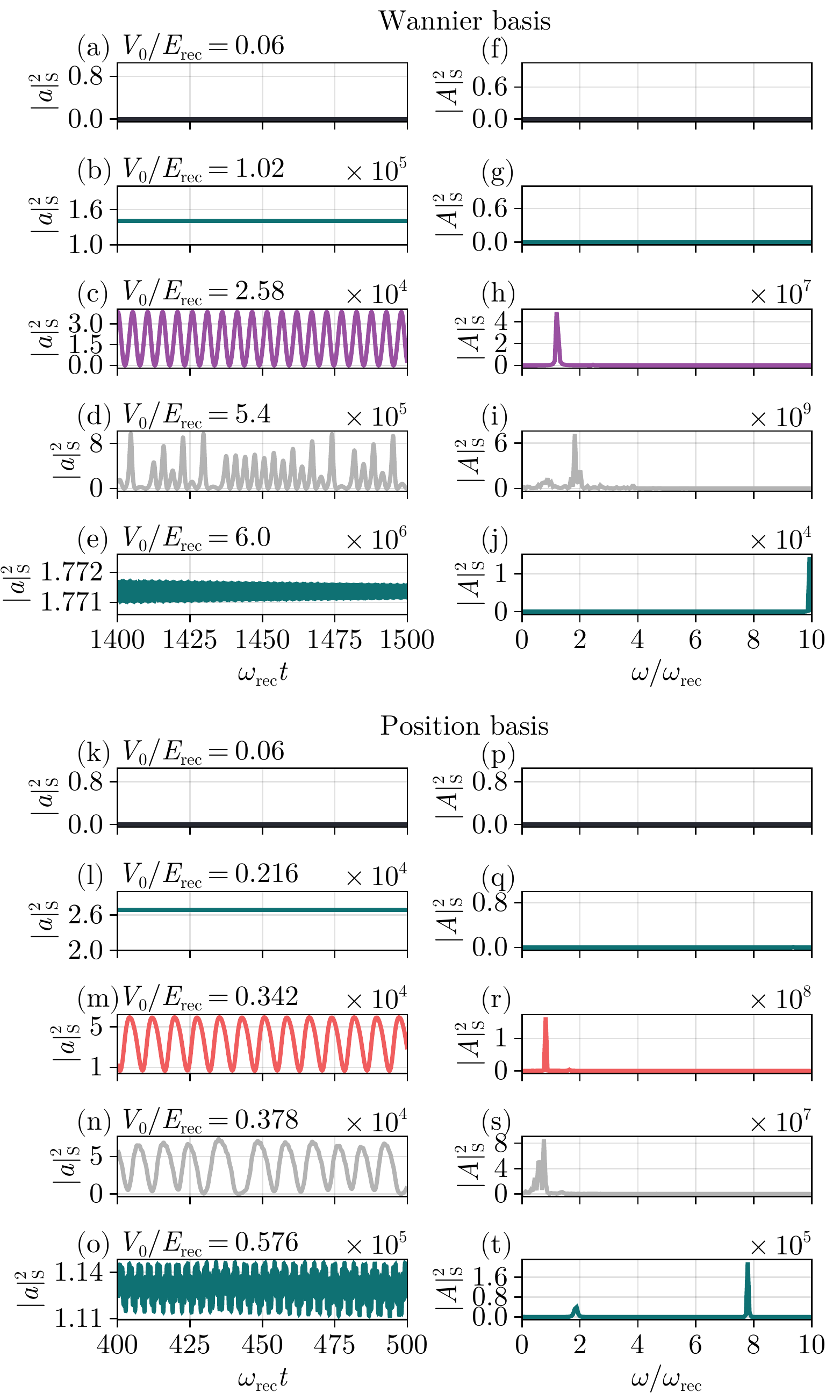}
    \caption{Dynamics of the (a)--(e), (k)--(o) long-time cavity-photon occupation $\abs{a}^2_\text{S}$, and (f)--(j), (p)--(t) their power spectra $\abs{A}^2_\text{S}$ for varying pump strength, according to the (a)--(j) Wannier, and (k)--(t) position bases. These correspond to points in Fig.~\ref{fig:dynamics}(c) and \ref{fig:dynamics}(d).}
    \label{fig:steady-state}
\end{figure*}

\section{STEADY-STATE DYNAMICS}\label{sec:finite_integration}

We show in Fig.~\ref{fig:steady-state} the exemplary long-time cavity-photon occupation dynamics $\abs{a}^2_\text{S}$ and their power spectrum corresponding to the data presented in Figs.~\ref{fig:dynamics}(c) and \ref{fig:dynamics}(d). As seen in Fig.~\ref{fig:steady-state}, the long-time cavity-photon occupation approaches zero for the SF phase. For the DW phase, while its $\abs{a}^2_\text{S}$ is generally nonzero and standard deviation is small, we sometimes observe transient oscillations for strong pump as depicted in Fig.~\ref{fig:steady-state}(t). To account for this, we set thresholds for the standard deviation of the cavity-photon occupation number, $\sigma(\abs{a}^2_\text{S})$, and spectral entropy $S$ for a response to be identified as a DW phase. Nevertheless, the transient oscillations in the DW phase are faster and have smaller amplitudes compared to those in the LC phase; see Figs.~\ref{fig:steady-state}(m) and \ref{fig:steady-state}(r) for LC and Figs.~\ref{fig:steady-state}(o) and \ref{fig:steady-state}(t) for DW. The amplitude of oscillation of the cavity-photon occupation is smaller in the DW phase at strong pump compared to that in the LC phase. There are also fewer peaks in the power spectrum of $\abs{a}^2_\text{S}$ in the DW phase compared to the LC phases.

From Fig.~\ref{fig:steady-state}, we see that the amplitude of oscillation of the cavity-photon occupation in the ID phase is greater than that in the LC phase. We can also observe that the $\abs{A}^2_\text{S}$ in the ID phase has more peaks than in the LC phase. This suggests that $\sigma(\abs{a}^2_\text{S})$ and $S$ is larger in the ID phase than in the LC phase, and motivates the use of threshold values for distinguishing the various dynamical phases in the main text.
    \section{DECORRELATOR}\label{sec:decorrelator}

The decorrelator $d$ is another order parameter that can be used to differentiate between the LC and ID phases. It has been used to benchmark the onset of chaos in discrete \cite{pizzi-2021} and dissipative time crystals \cite{cosme-2023,cosme-2025}.

To calculate the decorrelator, we first obtain two long-time solutions of the cavity-photon occupation $\abs{a}^2_1(t)$ and $\abs{a}^2_2(t)$ for two different initial states differing by $10^{-5}$ with respect to the initial values of the cavity-photon $a$ and bosonic $b$ occupation. The two solutions are then aligned corresponding to the appropriate first peak of the photon dynamics for each trajectory, over a time window of $60/\omega_\text{rec}$. Finally, the decorrelator is calculated using
\begin{equation}
    d=\frac{1}{N_\text{steps}}\sum_i\abs{\abs{a}^2_1(t_i^{(1)})-\abs{a}^2_2(t_i^{(2)})},
\end{equation}
where $N_\text{steps}$ is the number of recorded time steps in the time window, and $t^{(j)}$ is the $t^{(j)}$-th time adjusted after alignment of the two solutions.

\begin{figure}[ht!]
    \centering
    \includegraphics[width=\columnwidth]{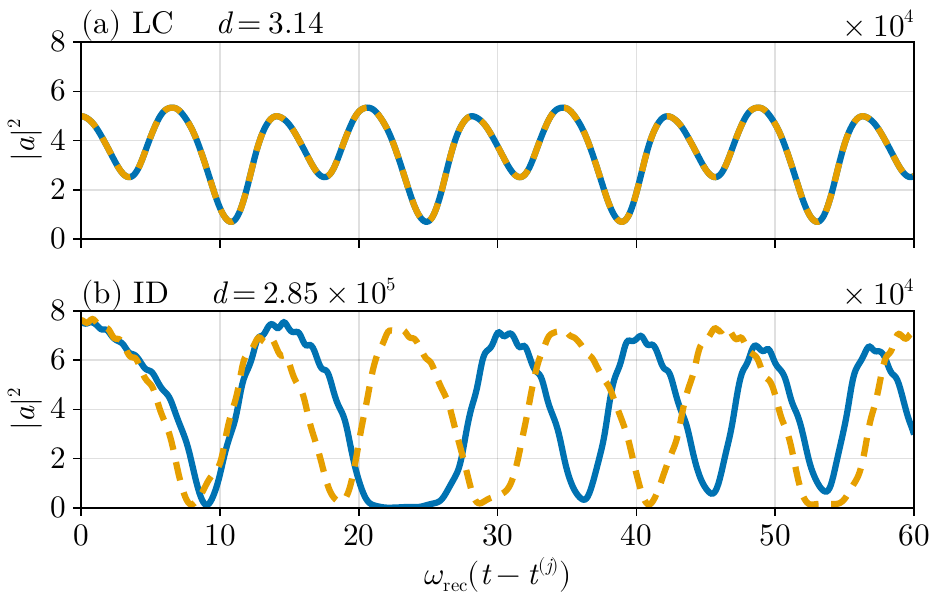}
    \caption{Aligned long-time cavity-photon occupation dynamics corresponding to the (a) LC and (b) ID phase, shown together with the corresponding decorrelator values. The lines result from slightly different initial states.}
    \label{fig:decorrelator-trajectory}
\end{figure}

In Fig.~\ref{fig:decorrelator-trajectory}, we show the aligned long-time cavity-photon occupation $\abs{a}^2$ dynamics in the LC and ID phases. Note that in the LC phase, both trajectories follow almost the same dynamics, which is in contrast to the dynamics in the ID phase. Also, the decorrelator in the ID phase is much greater than that in the LC phase. Since the cavity-photon occupation oscillates at a well-defined frequency in the LC phase, its long-time dynamics will not be affected by perturbations to the initial state resulting in a small $d$. On the other hand, since cavity-photon occupation behaves irregularly in the ID phase, then perturbations to the initial state will greatly affect the resulting long-time dynamics resulting in a very large $d$. Hence, the decorrelator can unambiguously distinguish between the LC and ID phases.

\begin{figure}[ht!]
    \centering
    \includegraphics[width=\columnwidth]{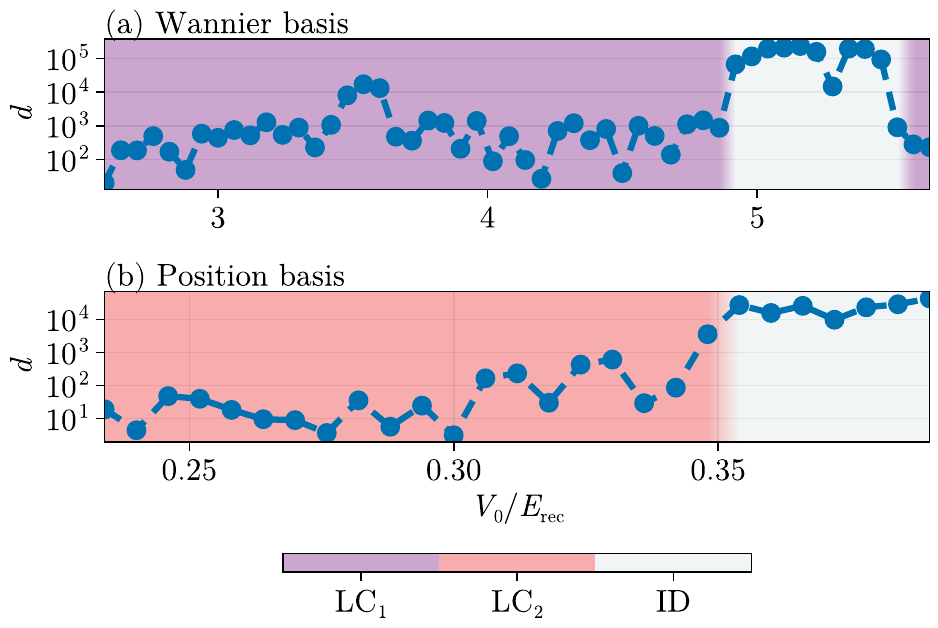}
    \caption{Decorrelator $d$ values corresponding to the LC-ID regions in Figs.~\ref{fig:dynamics}(c) and \ref{fig:dynamics}(d).}
    \label{fig:decorrelator-value}
\end{figure}

In Fig.~\ref{fig:decorrelator-value}, we now show the decorrelator for varying $V_0/E_\text{rec}$ within the LC-ID regions in Figs.~\ref{fig:dynamics}(c) and \ref{fig:dynamics}(d), corresponding to long-time cavity photon dynamics in the Wannier and position bases respectively. Notice that the decorrelator in the LC phase is indeed smaller compared to that in the ID phase, and follows a similar trend to the spectral entropy shown in Figs.~\ref{fig:dynamics}(c) and \ref{fig:dynamics}(d). Since the behavior of the decorrelator in the LC-ID region is similar to the behavior of the spectral entropy in the same region, then our choice of spectral entropy thresholds can also be used to distinguish between the LC and ID phases.
    \section{CONVERGENCE OF TWA RESULTS}\label{sec:twa_convergence}

We investigate the behavior of the long-time average of the largest natural occupation $\overline{n^\text{NO}_1}$ as a function of the number of trajectories used in our TWA simulations. To this end, we simulate an ensemble of 100 trajectories corresponding to the solutions of Eqs.~\eqref{eq:sec02-08} and \eqref{eq:sec02-09} in the presence of both initial quantum and stochastic noise.  We discretize one cavity wavelength to $M=16$ points, and considered $\Delta_\text{c}/E_\text{rec}=-8.18$ and $V_0/E_\text{rec}=6$. The long-time average of the largest natural occupation was calculated from the last $100/\omega_\text{rec}$ of the simulation.

\begin{figure}[!ht]
    \centering
    \includegraphics[width=\columnwidth]{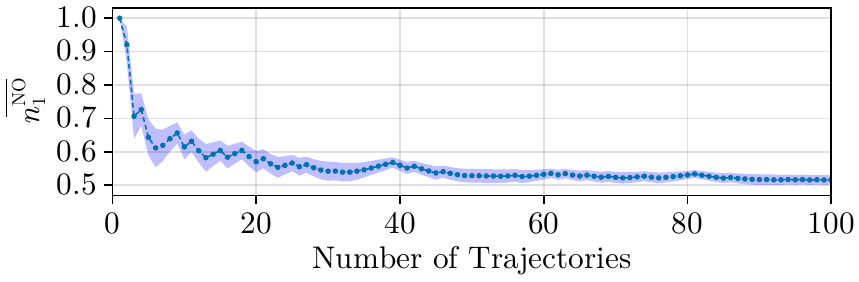}
    \caption{Long-time average of the largest natural occupation $\overline{n^\text{NO}_1}$ in the position basis as a function of the number of trajectories. The blue outline marks the standard deviation of $\overline{n^\text{NO}_1}$.}
    \label{fig:twa_convergence}
\end{figure}

We show in Fig.~\ref{fig:twa_convergence} the long-time average of the largest natural occupation as a function of the number of trajectories. The light-blue outline marks the standard deviation of $\overline{n^\text{NO}_1}$. As the number of trajectories is increased, the long-time average approaches a constant value, while its standard deviation decreases. This provides a benchmark for choosing the minimum number of trajectories to calculate the natural occupations and orbitals. In the main text, we have used 50 trajectories.
    \section{NATURAL ORBITALS}

\begin{figure}[ht!]
    \centering
    \includegraphics[width=\columnwidth]{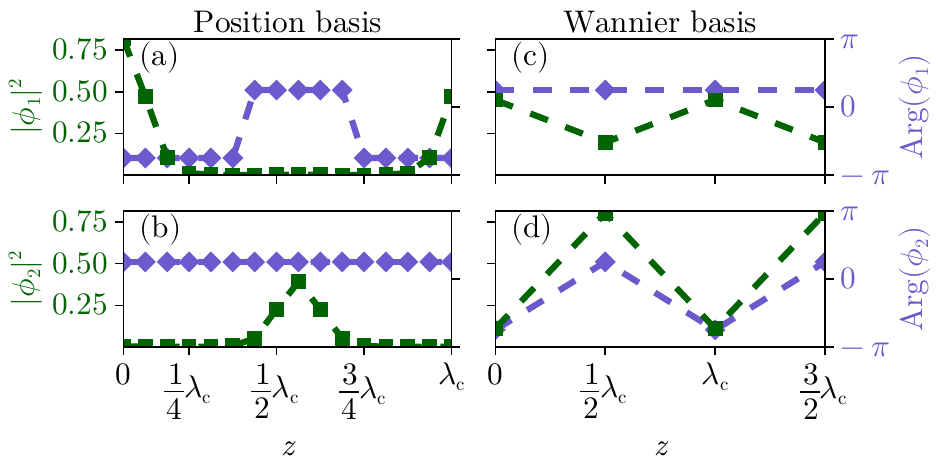}
    \caption{The square modulus and argument of the two lowest quantum states of the system with $U_\text{a}=0$ at steady state in the (a),(b) in the position and (c),(d) Wannier basis. These correspond to $V_0/E_\text{rec}=5$ of Figs.~\ref{fig:no_steady-state}(b) and \ref{fig:no_steady-state}(c). The ensemble average was taken over 120 trajectories for better convergence.}
    \label{fig:nat_orb}
\end{figure}

It is insightful to analyze the natural orbitals in the regime when two natural occupations have almost the same value $\sim0.5$. In Fig.~\ref{fig:nat_orb}, we show $\abs{\phi_i}^2$ and $\Arg(\phi_i)$ of the natural orbitals corresponding to the two largest natural occupations in both bases for pump intensity $V_0/E_\text{rec}=5$. The natural orbitals exhibit density patterns akin to the two sublattices or symmetry broken states of the DW phase. In the position basis, Figs.~\ref{fig:nat_orb}(a) and \ref{fig:nat_orb}(b), atoms could occupy either the outer edge or center of the cavity wavelength. On the other hand, atoms occupy either the even or odd lattice sites in the Wannier basis [see Figs.~\ref{fig:nat_orb}(c) and \ref{fig:nat_orb}(d)]. Note that the two highest occupied natural orbitals are degenerate in that they have almost the same natural occupation of $0.5$. This explains the apparent contradiction in the $\Arg(\phi_i)$ when comparing the Wannier and position bases since the labels $\phi_1$ and $\phi_2$ are interchangeable in this case.

    \section{NATURAL OCCUPATIONS FOR $U_\text{a}>0$}\label{sec:all_no_with_interaction}

\begin{figure}[ht!]
    \centering
    \includegraphics[width=\columnwidth]{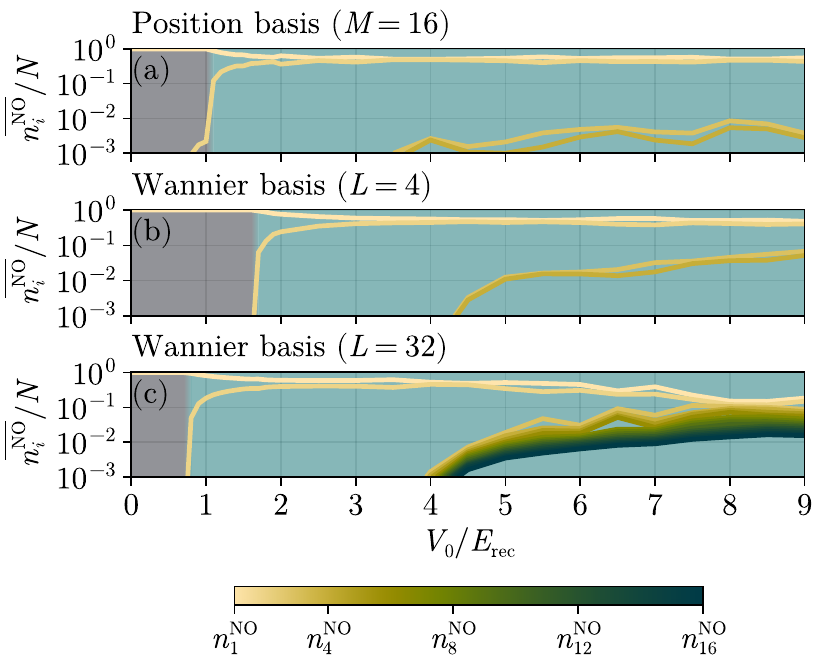}
    \caption{Long-time average of the sixteen largest natural occupations of bosons with contact interaction $U_\text{a}/E_\text{rec}=10^{-4}$ at steady state $\overline{n^\text{NO}_i}$ obtained using the (a) position basis, and Wannier basis with $\Delta_\text{c}/E_\text{rec}\approx-12.2$, (b) $L=4$ and (c) $L=32$ lattice sites}
    \label{fig:no_posU_all}
\end{figure}

In Fig.~\ref{fig:no_posU_all}, we show the long-time average of the sixteen largest natural occupations of bosons with contact interaction $U_\text{a}>0$ using a logarithmic scale. For large pump strengths, the Wannier basis predicts that at least 16 natural orbitals become significantly occupied, which is in contrast to the position basis prediction of only two to four natural orbitals. Since the position basis here is limited to $L_\text{sys}=\lambda_\text{c}$, our results suggest that larger system sizes are required to capture the correct fragmentation dynamics when $U_\text{a}>0$.

\begin{figure}[ht!]
    \centering
    \includegraphics[width=\columnwidth]{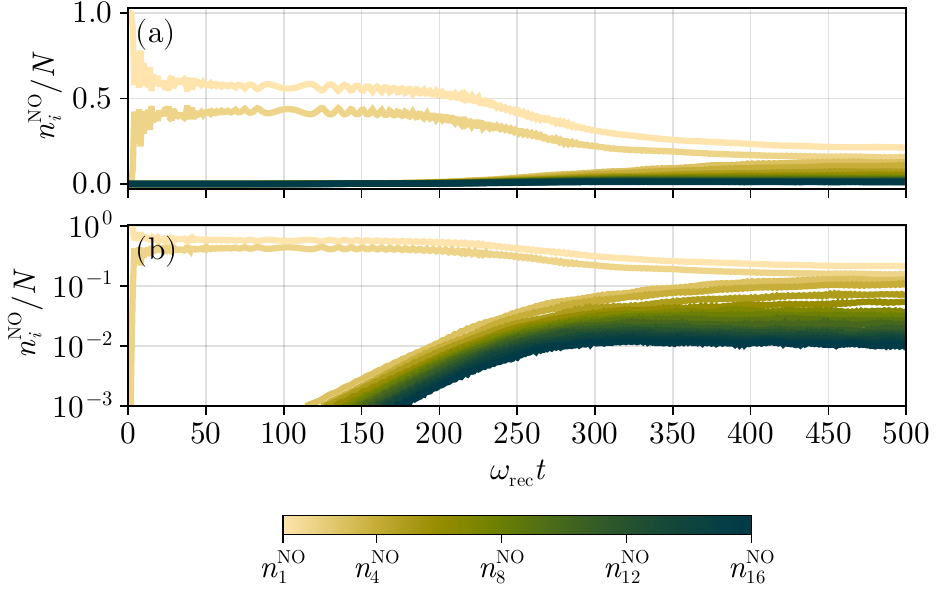}
    \caption{Dynamics of the natural occupations of the RSPDM in the Wannier basis for $L=32$ lattice sites, with contact interaction strength $U_\text{a}/E_\text{rec}=10^{-4}$ using (a) linear and (b) logarithmic scale.}
    \label{fig:prethermal}
\end{figure}

In Fig.~\ref{fig:prethermal}, we plot the natural occupation dynamics accompanying the simulations made for Figs.~\ref{fig:wig_dist}(c) and \ref{fig:wig_dist}(f). For time $\omega_\text{rec}t<500$, we have a prethermal-like state within $50\leq\omega_\text{rec}t\leq150$, wherein the system appears to be some fragmented state of two natural orbitals. However, eventually, the system relaxes to the true steady-state where coherence is almost absent as inferred from the high degree of fragmentation.

    \section{PHOTON-MEDIATED ATOM-ATOM INTERACTIONS}\label{sec:large_system_no_interaction}

\begin{figure}[ht!]
    \centering
    \includegraphics[width=\columnwidth]{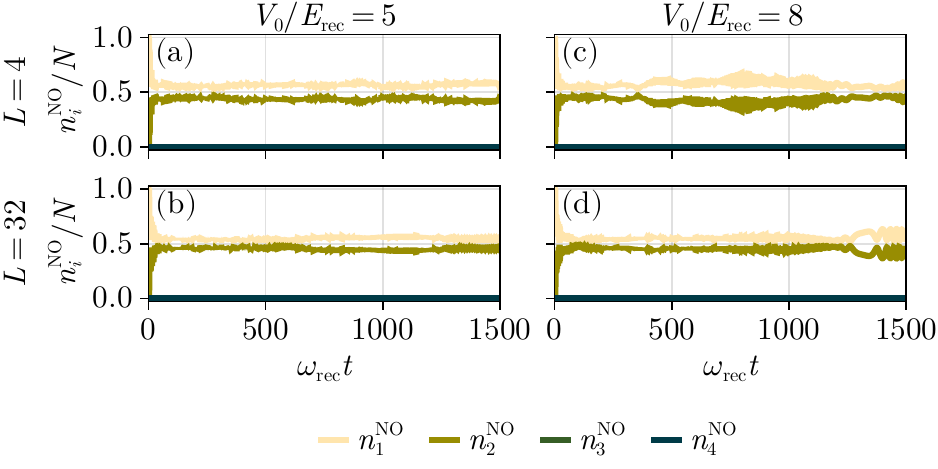}
    \caption{Dynamics of the four largest natural occupations of the RSPDM in the Wannier basis for (a),(b) $V_0/E_\text{rec}=5$ and (c),(d) $V_0/E_\text{rec}=8$ with $U_\text{a}=0$, simulated using (a),(c) $L=4$, and (b),(d) $L=32$ lattice sites.}
    \label{fig:no_int_comparison}
\end{figure}

In Appendix \ref{sec:finite_size}, we showed that $L=4$ sites are sufficient for the dynamics to converge for $U_\text{a}=0$, using the Wannier basis. Still, it is interesting to observe how the fragmentation dynamics behaves for larger system sizes. In Fig.~\ref{fig:no_int_comparison}, we present the dynamics of the four largest natural occupations for $L=4$ and $L=32$ using the same parameters in Appendix \ref{sec:finite_size}. To this end, we obtain the quantum dynamics for $U_\text{a}=0$ using the same parameters in Appendix \ref{sec:finite_size}, for pump intensity $V_0/E_\text{rec}=5$. Note that only the two largest natural occupations are macroscopically occupied for $U_\text{a}=0$ for both small ($L=4$) and large ($L=32$) system sizes. This is consistent with the results of our finite-size scaling analyses in Appendix \ref{sec:finite_size} which show that $L=4$ sites are sufficient to describe the dynamics in the absence of contact interaction.
    \section{PHASE FLUCTUATIONS BETWEEN ADJACENT SITES} \label{sec:phase_fluctuations}

\begin{figure}[ht!]
    \centering
    \includegraphics[width=\columnwidth]{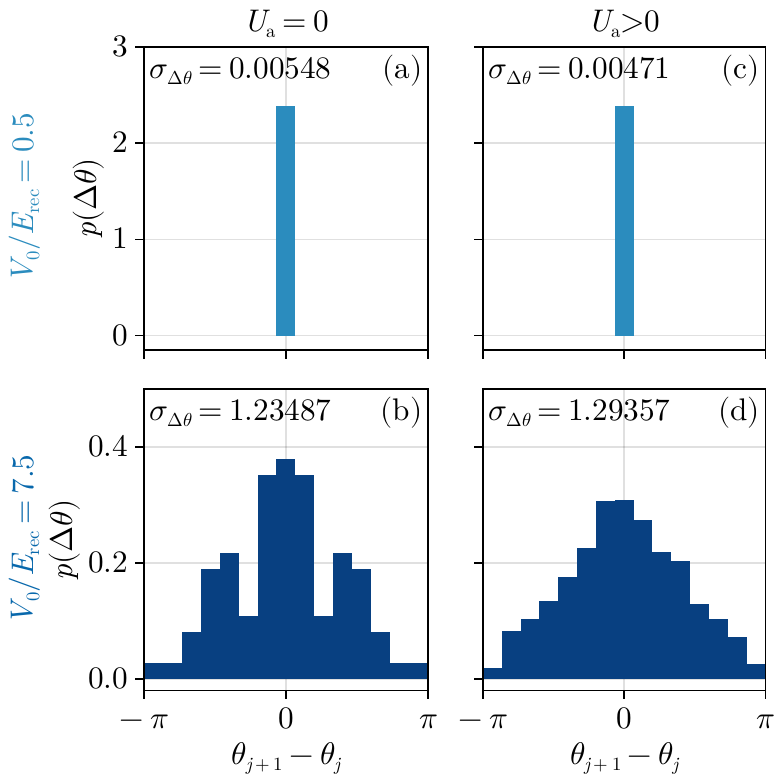}
    \caption{Steady-state distribution of the phase difference of the boson field between adjacent sites $p(\Delta\theta)$ for pump intensities $V_0$ as indicated. The corresponding standard deviation $\sigma_{\Delta\theta}$ is also shown for (a),(b) $U_\text{a}=0$ and (c),(d) $U_\text{a}>0$.}
    \label{fig:phase_fluctuations}
\end{figure}

In Fig.~\ref{fig:wig_dist}, we showed the steady-state Wigner distributions in the first site corresponding to one of the symmetry broken states in the DW phase. In addition to this, it is also interesting to investigate the phase differences between adjacent sites to characterize the overall coherence of the atoms across the lattice. In the following, we present the distribution of the phase difference of the boson field $\psi_j$ between adjacent sites for one of the symmetry broken states in the DW phase, corresponding to trajectories with $\Real(a)<0$ for the steady-state. Using the same parameters as in Sec.~\ref{sec:photon_atom_fragmentation}, we calculate the phase of the boson field at site $j$ using $\theta_j=\arctan[\Imag(\psi_j)/\Real(\psi_j)]$, where $\Imag(\psi_j)$ is the imaginary part of the boson field at site $j$. Using this, the phase difference between adjacent sites is obtained as:
\begin{equation}
    \Delta\theta = \theta_{j+1}-\theta_{j},
\end{equation}
where due to periodic boundary conditions, for a system modeled using $L$ sites, $\theta_{L+1}=\theta_1$. We also calculated the standard deviation of the phase difference $\sigma_{\Delta\theta}$ using
\begin{equation}
    \sigma_{\Delta\theta}=\sqrt{\sum_j(\Delta\theta_j-\overline{\Delta\theta})^2/N_{\Delta\theta}},
\end{equation}
where $\overline{\Delta\theta}$ is the average value of the phase difference, and $N_{\Delta\theta}$ is the number of recorded phase differences.

In Fig.~\ref{fig:phase_fluctuations}, we show the steady-state distribution of the phase differences between adjacent sites $p(\Delta\theta)$. The histograms are normalized so that $\int_{-\pi}^{\pi}p(\Delta\theta)\,\mathrm{d}(\Delta\theta)=1$, with bins obtained by dividing the range $[-\pi,\pi)$ to 15 uniform intervals. In the SF phase, Figs.~\ref{fig:phase_fluctuations}(a) and \ref{fig:phase_fluctuations}(c), we see that the distribution peaks at $\theta_{j+1}-\theta_j=0$, with a standard deviation $\sim10^{-3}$. This shows that atoms in the SF phase have phase coherence, indicative of a coherent state as expected of the SF phase \cite{olsen-2009}. In Figs.~\ref{fig:phase_fluctuations}(b) and \ref{fig:phase_fluctuations}(d), we show $p(\Delta\theta)$ in the DW phase for $V_0/E_\text{rec}=7.5$. We observe that the distribution of phase differences is now broadened, corroborating the loss of coherence inferred from Fig.~\ref{fig:wig_dist}.

    \bibliography{aps_references}
\end{document}